\DeclareSymbolFont{AMSb}{U}{msb}{m}{n}
\documentclass[11pt,a4paper,reqno,noamsfonts]{amsart}
\makeatletter 
\newcommand{\mylabel}[2]{#2\def\@currentlabel{#2}\label{#1}}
\renewcommand\@biblabel[1]{#1.}
\makeatother
\usepackage[english]{babel}
\usepackage[dvipsnames]{xcolor}
\usepackage{graphicx,pifont} 
\usepackage[utopia]{mathdesign}
\usepackage{bbm}
\usepackage[scr,scaled=1.0]{rsfso}
\usepackage{tikz-cd}

 \usepackage[pdftex,
                paper=a4paper,
                portrait=true,
                textwidth=160mm,
                textheight=247mm,
                tmargin=2.5cm,
                marginratio=1:1]{geometry}

\usepackage[utf8]{inputenc}
\usepackage{braket,caption,comment,mathtools,stmaryrd}
\usepackage[usestackEOL]{stackengine}
\usepackage{multirow,booktabs,microtype,relsize}
\usepackage[colorlinks,bookmarks]{hyperref} %
      \hypersetup{colorlinks,%
            citecolor=olive,%
            filecolor=black,%
            linkcolor=teal,%
            urlcolor=green}
      \numberwithin{equation}{section}
\usepackage[capitalise]{cleveref}

\def\id{\mathrm{id}}

\def\pt{\mathrm{pt}}

\newcommand{\ud}{\mathrm{d}\mkern0mu}
\newcommand{\gh}{\mathrm{gh}\mkern0mu}
\newcommand{\uud}{\mathbf{d}\mkern0mu}

\newcommand{\PP}{\ensuremath{\mathbbmss{P}}} 
 
\newcommand{\RR}{\ensuremath{\mathbbmss{R}}} 
\newcommand{\ZZ}{\ensuremath{\mathbbmss{Z}}} 
\newcommand{\HH}{\ensuremath{\mathbbmss{H}}} 

\newcommand{\Acal}{\mathscr{A}}
\newcommand{\Bcal}{\mathscr{B}}
\newcommand{\Ccal}{\mathscr{C}}
\newcommand{\Dcal}{\mathscr{D}}

\newcommand{\Gcal}{\mathscr{G}}

\newcommand{\Mcal}{\mathscr{M}}
\newcommand{\Ncal}{\mathscr{N}}
\newcommand{\Ocal}{\mathscr{O}}
\newcommand{\Pcal}{\mathscr{P}}
\newcommand{\Scal}{\mathscr{S}}

\newcommand{\Wcal}{\mathscr{W}}
\newcommand{\Xcal}{\mathscr{X}}

\newcommand{\gfrak}{\mathfrak{g}}

\newcommand{\GL}{\operatorname{GL}}
\newcommand{\SO}{\operatorname{SO}}
\newcommand{\so}{\operatorname{\mathfrak{so}}}

\newcommand{\Spin}{\operatorname{Spin}}
\newcommand{\Ad}{\operatorname{Ad}}
\newcommand{\ad}{\operatorname{ad}}
\newcommand{\im}{\operatorname{im}}
\renewcommand{\ker}{\operatorname{ker}}

\newcommand{\tr}{\operatorname{tr}}
\newcommand{\Lie}{\operatorname{Lie}}
\newcommand{\Pol}{\operatorname{Pol}}

\newcommand{\astDelta}{\operatorname{\ast\Delta}\!}

\def\Phibm{\boldsymbol{\Phi}}
\def\Wbm{\boldsymbol{W}}
\def\Ibm{\boldsymbol{I}}
\def\Abm{\boldsymbol{A}}
\def\cbm{\boldsymbol{c}}
\def\phibm{\boldsymbol{\phi}}
\def\rhobm{\boldsymbol{\rho}}

\def\sigmabm{\boldsymbol{\sigma}}

\newcommand{\lbbar}{\{\kern-0.76ex\{}
\newcommand{\rbbar}{\}\kern-0.76ex\}}

\newcommand*{\sbullet}{\raisebox{0.1ex}{\scalebox{0.6}{$\bullet$}}}

\title{A topological quantum field theory for $\Spin(7)$-instantons}

\author{Rafael Herrera} 
\address{Centro de Investigaci\'{o}n en Matem\'{a}ticas \\ A. P. 402, 36000 Guanajuato \\ Gto. \\ M\'{e}xico}
\email{rherrera@cimat.mx}

\author{Sergio A. Holgu\'in Cardona} 
\address{Secihti Research Fellow--Instituto de Matem\'aticas\\ Universidad Nacional Aut\'{o}noma de M\'{e}xico \\ Leon 2 altos, Col. centro, 68000 Oaxaca \\ Oax. \\ M\'exico}
\email{sholguin@im.unam.mx}

\author{Alexander Quintero V\'{e}lez} 
\address{Departamento de Matem\'{a}ticas\\ Universidad Nacional de Colombia Sede Medell\'{i}n \\ Carrera 65 $\#$ 59A--110 \\ Medell\'{i}n \\ Colombia}
\email{aquinte2@unal.edu.co}

\begin{document}

\begin{abstract}
We construct a topological quantum field theory based on the moduli space of $\Spin(7)$-instantons on 8-dimensional manifolds. Using the Mathai-Quillen formalism, we derive the action of the theory in purely geometric terms, which coincides with prior results in the literature. We then reformulate the theory within the AKSZ formalism, obtaining a Batalin–Vilkovisky action that, after gauge fixing, matches our Mathai-Quillen construction while making the BRST symmetry explicit and providing a natural framework for classical observables. We also show that the Batalin–Vilkovisky action can be elegantly recast as a Chern–Simons type theory.
\end{abstract}

\maketitle

\section{Introduction}
Topological quantum field theories have attracted significant interest over nearly four decades. These theories can be broadly defined as those where the correlation functions of observables are topological invariants. From a mathematical perspective, they offer a valuable opportunity to discover new topological invariants within the framework of quantum field theory. Physically, they are interesting because they describe systems in which observables depend only on global, topological properties, rather than on local geometric features.

Perhaps one of the most well-known and influential examples, and one that helped establish this area of research, is what we now call Donaldson-Witten theory. Developed by Witten in \cite{witten1988topological}, this theory provides a field-theoretic framework for the Donaldson invariants, which are topological invariants of $4$-dimensional manifolds. This line of work has been the subject of several reviews over the years, including early ones such as \cite{birmingham1991topological} and \cite{marino2003introduction}, and a more recent account in \cite{moshayedi20224}. Soon after, an elegant approach by Atiyah and Jeffrey \cite{Atiyah-Jeffrey1990}, building on the work of Mathai and Quillen \cite{Mathai-Quillen1986} to construct ``Gaussian representatives'' of Thom forms for finite-dimensional vector bundles, revealed that the Donaldson-Witten theory could be understood as an infinite-dimensional generalization of this construction. In more detail, they showed that the partition function of the theory can be interpreted as the Euler number of the moduli space of anti-self-dual connections, or instantons, on the $4$-dimensional manifold. This perspective led to a number of further developments, with various topological quantum field theories being constructed from finite-dimensional moduli spaces using what has come to be known as the Mathai-Quillen formalism \cite{blau1993mathai,blau1993n,merkulov1994topological,labastida1995topological,Anguelova:2005cv}.

In the first part of this paper, we construct the topological quantum field theory associated with the moduli problem considered in \cite{Munoz-Shahbazi2020}. The moduli space in question is that of $\Spin(7)$-instantons on an $8$-dimensional manifold equipped with a $\Spin(7)$-structure. The resulting theory turns out to be one that was previously studied by Acharya, O'Loughlin, and Spence in \cite{A-O-S1997}, and by Baulieu, Kanno, and Singer in \cite{B-K-S1998}. However, in our approach, since we make use of the Mathai-Quillen formalism, the action of the topological quantum field theory is derived directly from the moduli space of $\Spin(7)$-instantons, and thus it can be formulated in purely geometric terms.

With the action of the quantum field theory at hand, the next natural step is to exhibit its BRST invariance and determine the classical observables. However, the Mathai-Quillen formalism does not provide a direct or systematic method for achieving this. One can, of course, emulate the approach taken in the Donaldson-Witten theory and attempt to interpret the BRST operator and the observables through an infinite-dimensional version of the Cartan model for equivariant cohomology on the field space. Our initial strategy will be to proceed along these lines, which, as we shall see, leads naturally to the observables previously proposed in \cite{A-O-S1997} and \cite{B-K-S1998}. Nevertheless, to cast the theory within a more powerful and unified framework, we will adopt the paradigm introduced by Alexandrov, Kontsevich, Schwarz, and Zaboronsky in \cite{AKSZ1997}, in what is now commonly referred to as the AKSZ formalism. 

The AKSZ method is a remarkably simple and elegant geometrical construction of solutions of the classical master equation in the Batalin-Vilkovisky formalism. It encodes them through geometrical data, which is compactly expressed in the language of graded geometry. The construction encompasses a wide range of examples including the Poisson sigma model \cite{Cattaneo-Felder2001,bonechi2009poisson,Bonechi:2011um}, the Courant sigma model \cite{Roytenberg2007}, the topological A- and B-models \cite{Park:2000au,Cattaneo:2009zx,Bonechi:2018icn,Kokenyesi:2018xgj}, topological membranes on $G_2$-manifolds \cite{Kokenyesi:2018ynq}, and the Rozansky–Witten theory \cite{qiu2009aksz}. Further applications of the AKSZ prescription can be found, for instance, in \cite{Ikeda:2010vz}, \cite{Fiorenza:2011jr} and \cite{Velez:2011ed}. Of particular relevance to our purposes are the references \cite{Ikeda:2011xr} and \cite{bonechi2020equivariant}, which were the first to employ the AKSZ formalism to derive the Donaldson-Witten theory. Related work on this approach can be found in \cite{bonechi2023towards}, \cite{moshayedi20224}, and \cite{Ben-Shahar:2024dju}.

In the second part of this paper, we discuss the AKSZ formulation of the topological quantum field theory associated with the moduli space of $\Spin(7)$-instantons that we have constructed. The master action is derived from a suitable choice of target graded manifold, and shown to coincide, after gauge-fixing, with the action obtained via the Mathai-Quillen formalism. Crucially, within this framework, the BRST symmetry is naturally built-in, which renders the analysis of classical observables both transparent and structurally clear. Finally, by carefully grouping the ``superfields’’ involved, we show that the Batalin–Vilkovisky action admits a reformulation as a Chern–Simons type theory. This perspective, originally introduced in \cite{bonechi2023towards} in the context of Donaldson–Witten theory and further developed in \cite{Ben-Shahar:2024dju} in connection with color-kinematics duality, offers a compelling and geometrically transparent formulation of the theory.

The organisation of this paper is as follows. Section~\ref{sec:2} provides an overview of the essential concepts related to $\Spin(7)$-structures, the Mathai-Quillen formalism, and the AKSZ construction for topological quantum field theories, while establishing the notation and terminology used in the rest of the paper. In section~\ref{sec:3}, we construct the topological quantum field theory that encodes the geometry of the moduli space of $\Spin(7)$-instantons, using the Mathai-Quillen formalism in the framework of Atiyah and Jeffrey, and establish that the theory coincides with the one outlined in \cite{A-O-S1997} and \cite{B-K-S1998}. Additionally, we examine the classical observables through the lens of the infinite-dimensional Cartan model for equivariant cohomology over the field space, showing that this approach recovers the types of observables discussed in those works. Finally, section~\ref{sec:4} presents the AKSZ formulation of the topological quantum field theory for $\Spin(7)$-instantons obtained in section~\ref{sec:3}, leading to a Batalin–Vilkovisky action that, upon gauge fixing, exactly matches the structure and content of the theory developed there. We also describe the classical observables of the theory, which originate from the cohomology of the target graded manifold, and show that the Batalin–Vilkovisky action can be recast as a Chern–Simons type theory, in the sense mentioned above.

\subsection*{Acknowledgements}
The authors would like to express their gratitude to the Centro de Investigación en Matemáticas (CIMAT) for its warm hospitality during the academic visits in which this work took shape. A.Q.V. thanks Francesco Bonechi for valuable discussions related to the results of this work.


\section{Preliminaries}\label{sec:2}
The aim of this section is to recall the relevant information concerning $\Spin(7)$-structures, the Mathai-Quillen formalism and the AKSZ construction of topological field theories. This will, among other things, establish the notation and terminology that will be used throughout the rest of the paper.

\subsection{$\Spin(7)$-structures}\label{sec:2.1}
In this subsection we provide the necessary background on $\Spin(7)$-structures, following the presentation of \cite{Munoz-Shahbazi2020}. The reader might see \cite{joyce2000compact} or \cite{Joyce2003} for a more thorough treatment.

Let $\RR^{8}$ have coordinates $x^{1},\dots, x^{8}$. Define a $4$-form $\Omega_0$ on $\RR^{8}$ by
\begin{equation}
\begin{aligned}
\Omega_0 &= \ud x^{1234} - \ud x^{1278} - \ud x^{1638} - \ud x^{1674} + \ud x^{1526} + \ud x^{1537} + \ud x^{1548} \\
&\quad \: + \ud x^{5678} - \ud x^{5634} - \ud x^{5274} - \ud x^{5238} + \ud x^{3748} + \ud x^{2648} + \ud x^{2637},
\end{aligned}
\end{equation}
where $\ud x^{\mu\nu\rho\sigma}$ stands for $\ud x^{\mu} \wedge \ud x^{\nu} \wedge \ud x^{\rho} \wedge \ud x^{\sigma}$. The subgroup of $\GL(8,\RR)$ preserving $\Omega_0$ is isomorphic to $\Spin(7)$, which is a simply-connected, compact, Lie group of dimension $21$, abstractly isomorphic to a double cover of $\SO(7)$. It also preserves the standard orientation of $\RR^{8}$ and the euclidean metric $g_{0} = (\ud x^{1})^2 + \cdots + (\ud x^{8})^2$, so that it can realised as a subgroup of $\SO(8)$. Moreover, it is easy to check that $\Omega_0 = \ast \Omega_0$, where $\ast$ is the Hodge star operator defined by $g_0$.

We shall use some representation theory of $\Spin(7)$. The standard representation is $V =  \RR^{8}$ with the action given by the inclusion of $\Spin(7)$ into $\GL(8,\RR)$. Write $\Lambda^2 = \Lambda^2 V$ for the two-fold exterior power of $V$. Then the decomposition of $\Lambda^2$ into its irreducible components under the action of $\Spin(7)$ is given by
\begin{equation}
\Lambda^2 = \Lambda^2_{7} \oplus \Lambda^2_{21},
\end{equation}
where the subindices indicate the corresponding dimension. These irreducible components are explicitly described as follows:
\begin{equation}
\begin{aligned}
\Lambda^2_{7} &= \{\alpha \in \Lambda^2 \mid \ast (\alpha \wedge \Omega) = 3 \alpha\}, \\
\Lambda^2_{21} &=  \{\alpha \in \Lambda^2 \mid \ast (\alpha \wedge \Omega) = - \alpha\}.
\end{aligned}
\end{equation}
From this, we can write down the formulas for the projections $\pi_{7} \colon \Lambda^2 \to \Lambda^2_{7}$ and $\pi_{21} \colon \Lambda^2 \to \Lambda^2_{21}$ as
\begin{equation}\label{eq:2.4}
\begin{aligned}
\pi_{7}(\alpha) &= \tfrac{1}{4} \left(\alpha + \ast (\alpha \wedge \Omega) \right), \\
\pi_{21}(\alpha) &= \tfrac{1}{4} \left(3\alpha - \ast (\alpha \wedge \Omega) \right).
\end{aligned}
\end{equation}
Note in particular that these projections are complementary, that is, they satisfy $\pi_{7} + \pi_{21} = \id_{\Lambda^2}$. They are also mutually orthogonal, in the sense that $\pi_{7}\circ \pi_{21} = \pi_{21}\circ \pi_{7} = 0$. 

Next we shall discuss $\Spin(7)$-structures. Let $M$ be an $8$-dimensional manifold. A $\Spin(7)$-\emph{structure} on $M$ is the choice of a $4$-form $\Omega$ on $M$, such that each tangent space of $M$ admits an isomorphism with $\RR^{8}$ identifying $\Omega$ with $\Omega_0$. Given such a $\Spin(7)$-structure $\Omega$, we define its torsion as $W = \ud \Omega$. It should be noted that the $4$-form $\Omega$ induces a riemannian metric $g$ and an orientation on $M$ via the inclusion of $\Spin(7)$ into $\SO(8)$. It can be shown that the holonomy of $g$ is contained in $\Spin(7)$ precisely when $\nabla \Omega = 0$, where $\nabla$ is the Levi-Civita connection associated to $g$. By \cite{Fernandez1986}, this is equivalent to $\ud \Omega = 0$. In this case the $\Spin(7)$-structure is called \emph{integrable}. If $W = \ud \Omega \neq 0$ then we say that $\Omega$ is a non-integrable $\Spin(7)$-structure. 

Although explicit examples of closed $8$-manifolds with integrable $\Spin(7)$-structures are scarce \cite{Joyce1996}, examples illustrating $\Spin(7)$-structures can be obtained from closed $8$-dimensional manifolds admitting almost-quaternionic structures. In fact, it was proven in Proposition~3.3 of  \cite{Munoz-Shahbazi2020} that such manifolds always admit an $\Spin(7)$-structure (in general unrelated to the existent almost-quaternionic structure). As a particular case of this result, we really know that the $8$-dimensional quaternionic space $\PP^2(\HH)$ carries an $\Spin(7)$-structure. For more examples of manifolds admitting such structures, the reader can see \cite{fernandez1987new} and \cite{joyce1999new}. 

Let $M$ be an $8$-dimensional manifold with an $\Spin(7)$-structure $\Omega$. The action of $\Spin(7)$ on each tangent space of $M$ gives rise to an action of $\Spin(7)$ on the two-fold exterior power $\Lambda^2 T^* M$. Thus, as before, we get a decomposition of $\Lambda^2 T^* M$ into its irreducible components under the action of $\Spin(7)$, which we denote by $\Lambda^2_{7} T^* M$ and $\Lambda^2_{21} T^* M$. It follows at once from this, on setting $\Omega^2_{7}(M) = \Gamma(\Lambda^2_{7} T^* M)$ and  $\Omega^2_{21}(M) = \Gamma(\Lambda^2_{21} T^* M)$, that the space of $2$-forms on $M$ splits orthogonally as
\begin{equation}
\Omega^2(M) = \Omega^2_{7}(M) \oplus \Omega^2_{21}(M). 
\end{equation}
Accordingly, we write $\pi_7 \colon \Omega^2(M) \to \Omega^2_{7}(M)$ and $\pi_{21} \colon \Omega^2(M) \to \Omega^2_{21}(M)$ for the orthogonal projections onto the respective summands. They are determined by the same formulas as those in \eqref{eq:2.4}.

\subsection{The Mathai-Quillen formalism} \label{sec:2.2}
In this subsection we briefly discuss the Mathai-Quillen construction of a set of representatives of the Euler class for finite-dimensional vector bundles. We refer the reader to \cite{Wu2006} for an excellent review of this whole subject.

We start with some classical material. Recall that an oriented $2m$-dimensional real vector bundle $E$  over a manifold $M$ has an Euler class $e(E) \in H^{2m}(M,\ZZ)$. If the dimension of $M$ is $2m$, this class can be evaluated on the fundamental class $[M]$ of $M$ to give the Euler number $\chi(E) = e(E)[M]$. There are two quite different prescriptions for calculating $\chi(E)$. The first is topological in nature and instructs one to choose a section of $E$ with isolated zeros and count these zeros with signs. The second is differential geometric and represents $\chi(E)$ as the integral over $M$ of a $2m$-form constructed from the curvature of some connection on $E$. 

A more general formula, obtained by Mathai and Quillen \cite{Mathai-Quillen1986}, interpolates between these two classical prescriptions. Their construction involves what is known as the Thom class of $E$ or, more precisely, an explicit differential form representative of such class. The construction of this representative is best understood within the framework of equivariant cohomology. What this means is that one does not work directly on $E$, but that one realises $E$ as an associated vector bundle $P \times_{\rho} V$, where $P$ is a principal bundle with structure $G$ and $\rho \colon G \to \SO(V)$ is a $2m$-dimensional orthogonal representation of $G$, and works equivariantly on $P \times V$. To be more specific, it will be necessary to introduce some more notation. 

We denote by $\gfrak$ the Lie algebra of $G$ and by $\rho_{*} \colon \gfrak \to \so(V)$ the Lie algebra homomorphism induced by $\rho$. Choose a connection $\omega$ on $P$ with curvature $\Omega$. We also let $\{\chi^{1},\dots,\chi^{2m}\}$ be a fixed orthonormal basis for $V$ and $\{x_1,\dots,x_{2m}\}$ its dual basis of coordinate functions on $V$. Then the Mathai-Quillen form $U$ is the basic form on $P \times V$ which one can write formally as the horizontal projection of
\begin{equation}
(2\pi)^{-m} \mathrm{e}^{-\lVert x \rVert^2} \int \exp \left\{ i \chi^{\mathrm{T}} \ud x + \tfrac{1}{4}\chi^{\mathrm{T}}(\rho_*\Omega) \chi \right\} \Dcal\!\chi.
\end{equation}
Here $\lVert x \rVert^2 = x_1^2 + \cdots + x_{2m}^2$, $\chi$ is the column matrix $(\,\chi^1\: \cdots \:\chi^{2m}\,)^{\mathrm{T}}$, $\ud x$ is the column matrix $(\,\ud x_1\: \cdots \:\ud x_{2m}\,)^{\mathrm{T}}$ and the fermionic integral $\int \Dcal\chi$ means that we expand and take the coefficient of $\chi^1 \cdots \chi^{2m}$. To get a formula for the Euler class of the vector bundle $P \times_{\rho} V$ we must pick a section $s \colon M \to P \times_{\rho} V$ and pull $U$ back by $s$.  

Now our interest in the Mathai-Quillen formalism stems from the fact that Atiyah and Jeffrey \cite{Atiyah-Jeffrey1990} have shown how it can be adapted and formally applied to the case when the vector bundle is infinite-dimensional. We briefly sketch their construction, which works still in the finite-dimensional context. For this purpose, some extra assumptions on the various ingredients need to be added. First, we assume that $P$ is oriented and that the action of $G$ on $P$ preserves the orientation. Second, we assume that both $P$ and $\gfrak$ are provided with a $G$-invariant riemannian metric $\langle , \rangle$ and an $\ad$-invariant inner product $(,)$, respectively. Finally, we assume that the connection $\omega$ for $P$ is the one whose distribution of horizontal subspaces is the family of orthogonal complements with respect to $\langle , \rangle$ of the $G$-orbits in $P$.  

With these assumptions in place, we proceed to rewrite the Mathai-Quillen form $U$ following the Atiyah-Jeffrey point of view. So let $\Xcal(P)$ be the space of vector fields on $P$ and set $d = \dim G$. The action of $G$ on $P$ then defines a map $C \colon \gfrak \to \Xcal(P)$ that may be thought of as the ``infinitesimal'' action of $G$. Using the $G$-invariant riemannian metric on $P$ and the $\ad$-invariant inner product on $\gfrak$, we can consider the adjoint map $C^* \colon \Xcal(P)\to \gfrak$ and the map $R = C^* \circ C \colon \gfrak \to \gfrak$. In terms of these, we can express the connection on $P$ as $\omega = R^{-1} C^*$, while the curvature takes the form $\Omega = R^{-1} \ud C^*$ on horizontal vectors. Now, to enforce the horizontal projection we should have to integrate over the vertical degrees of freedom which amounts to an  integration over the Lie algebra $\gfrak$. Alternatively, we can  introduce a ``projection form'' which, besides of projecting on the horizontal direction, automatically involves the Chern-Weil homomorphism which substitutes the universal curvature by the actual curvature on $P$. The projection form also allows us to write the correlation functions on $M$ as integrals over $P$. Taking into account all these facts, and after some suitable manipulations, we obtain the following expression for the Mathai-Quillen form:
\begin{equation}
\begin{aligned}
U &=  (2 \pi)^{-d} (2\pi)^{-m} \mathrm{e}^{-\lVert x \rVert^2} \int \exp \left\{ i \chi^{\mathrm{T}} \ud x + \tfrac{1}{4}\chi^{\mathrm{T}}(\rho_*\phi) \chi  - i (\ud C^*,\bar{\phi}) \right. \\
& \phantom{=  (2 \pi)^{-d} (2\pi)^{-m} \mathrm{e}^{-\lVert u \rVert^2} \int \exp}  \left.\phantom{\tfrac{1}{4}} + i ( \phi, R \bar{\phi} ) + (C^*,\eta) \right\} \Dcal\!\eta  \, \Dcal\!\chi \, \ud \phi  \, \ud \bar{\phi}.
\end{aligned}
\end{equation}
Here in addition to the two fermionic integrations over $\eta$ and $\chi$ and the integrations over the Lie algebra  variables $\phi$ and $\bar{\phi}$, we also understand an integration over the fiber of $P \times V \to P \times_{\rho} V$. Again, pulling $U$ back to $M$ by a section $s \colon M \to P \times_{\rho} V$ gives a representative of the Euler class of $P \times_{\rho} V$ which, when integrated over $M$, gives the Euler number of $P \times_G  V$. However, we now note that pulling the $V$-factors of $U$ back by the $G$-equivariant map $S \colon P \to V$ naturally associated with $s$, gives a form on $P$ whose integral over $P$ is also the Euler number of $P \times_{\rho} V$. To put this formally, we use the standard trick of replacing differentials on $P$ by fermionic variables and integrating over them as well. With all this in mind the integral formula for the Euler number of $P \times_{\rho} V$ becomes
\begin{equation}\label{eq:2.8}
\begin{aligned}
&(2 \pi)^{-d} (2\pi)^{-m} \int \exp \left\{ - \lVert S(u)\rVert^2 + i \chi^{\mathrm{T}} \ud S_u(\psi) + \tfrac{1}{4}\chi^{\mathrm{T}}(\rho_*\phi) \chi  - i (\ud C^*_u(\psi,\psi),\bar{\phi}) \right. \\
& \phantom{(2 \pi)^{-d} (2\pi)^{-m} \int \exp}  \left.\phantom{\tfrac{1}{4}} + i ( \phi, R_u \bar{\phi} ) + (C^*_u \psi  ,\eta) \right\} \Dcal\!\psi \, \Dcal\!\eta \, \Dcal\!\chi \, \ud \phi \, \ud \bar{\phi}  \, \ud u.
\end{aligned}
\end{equation}
In this expression, for the sake of clarity, we have indicated explicitly the variables on which each term depends. We will refer to it as the Atiyah-Jeffrey formula for the Euler number of $P \times_{\rho} V$.  

\subsection{The AKSZ formalism}\label{sec:2.3}
We shall now proceed to discuss the AKSZ formalism, which is an important framework for the construction of topological field theories of various kinds, internal to the Batalin-Vilkovisky formalism. For a detailed description, including the mathematical background and implementations of the construction, the reader should consult the original paper \cite{AKSZ1997} and the later expositions in \cite{Cattaneo-Felder2001} and \cite{Roytenberg2007}. In addition, for background on graded geometry--whose language will be extensively used in what follows--and its relation to the Batalin-Vilkovisky formalism, we refer the reader to \cite{albert2010batalin}, and also to \cite{mnev2019lectures} for a general exposition of the Batalin-Vilkovisky formalism and its applications to a variety of topological gauge theories.

We begin by introducing the basic geometric ingredients. Firstly, a source manifold must be chosen, and this is taken to be a graded manifold $\Mcal$ endowed with a homological vector field $D$ and a measure $\mu$ of degree $-n$ for some positive integer $n$ such that the measure is invariant under $D$. In practice we will consider $\Mcal$ to be a shifted tangent bundle $T[1]M$ for some oriented $n$-manifold $M$, with $D$ the de Rham vector field and $\mu$ the canonical measure. Choosing local coordinates $\{x^{i}\}$ on $M$ together with their odd counterparts $\{\theta^{i}\}$, we can write $D = \theta^{i} \partial / \partial x^{i}$ and $\mu = \ud^{n}\theta \, \ud^{n} x$, in which we have put $\ud^{n} \theta = \Dcal\! \theta^{1} \cdots  \Dcal\! \theta^{n}$ and $\ud^{n} x = \ud x^{1} \wedge \cdots \wedge \ud x^{n}$. The second input is the target, and this is taken to be a graded manifold $\Ncal$ endowed with a symplectic form $\omega$ of degree $n-1$ and a homological vector field $Q$ preserving $\omega$. We require that $Q$ is hamiltonian, that is, there exists a function $\Theta$ of degree $n$ such that $Q = \{\Theta, \cdot \}$, where here $\{,\}$ is the odd Poisson bracket induced by $\omega$. This implies that $\Theta$ satisfies the following Maurer-Cartan equation:
\begin{equation}
\{\Theta,\Theta\} = 0. 
\end{equation}
In what follows, we shall assume that $\Ncal$ is nonnegatively graded. 

In the AKSZ formalism, a field configuration is defined by what is called a \emph{superfield}, that is, a map $\Phibm \colon T[1]M \to \Ncal$.\footnote{The phrase ``let $\Phibm$ be a map from $T[1]M$ to $\Ncal$'' is an abuse of language for ``let $\Pcal$ be a graded manifold and let $\Phibm$ be a map from $T[1]M \times \Pcal$ to $\Ncal$. See section 2.3 of \cite{Roytenberg2007} for a fuller explanation.} The space of all such superfields inherits a grading from $\Ncal$, which is commonly referred to as \emph{ghost number}. Picking local coordinates $\{u^{a}\}$ on $\Ncal$, a superfield $\Phibm$ can be described in terms of its components $\Phibm^{a} = u^{a} \circ \Phibm$. Each of these, in turn, can be represented using the local coordinates on $T[1]M$ as
\begin{equation}
\Phibm^{a} =\sum_{k=0}^{n} \Phi^{(k)a}, 
\end{equation}
where we have written $\Phi^{(k)a}$ for $\frac{1}{k!} \Phi_{i_1\cdots i_k}^{(k)a} \theta^{i_1} \cdots \theta^{i_k}$. The coefficients $\Phi_{i_1\cdots i_k}^{(k)a}$ are the components of a $k$-form on $M$ with values in $\Phibm^* \Ncal$. They are assigned a ghost number such that $\Phibm^{a}$ has ghost number equal to the degree of $u^{a}$.  In the physics literature, components of non-negative ghost number are called \emph{fields}, while those of negative ghost number are called \emph{antifields}. Among the fields one further distinguish between the \emph{classical fields}, of ghost number zero, and the \emph{ghost fields}, of positive ghost number. 

The space of superfields $\Phibm$ is endowed with an odd symplectic form $\Omega$ of ghost number $-1$, which is obtained from the symplectic form $\omega$ on $\Ncal$ upon integration over $T[1]M$. Writing things out in local coordinates $\{u^{a}\}$ on $\Ncal$, we have
\begin{equation}\label{eq:2.11}
\Omega =\tfrac{1}{2} \int_{T[1]M} \uud \Phibm^{a} \omega_{ab} \uud \Phibm^{b} \:  \ud^{n} \theta \, \ud^{n} x. 
\end{equation}
Here the symbol $\uud$ stands for the exterior derivative on the space of superfields and $\omega_{ab}$ are the components of $\omega$  in the coordinarte system $\{u^{a}\}$. Just as in the usual case, one can define odd Poisson bracket $(,)$ acting on functionals of the superfields. These Poisson brackets, which are the Batalin-Vilkovisky antibrackets, obey a graded Jacobi identity. The space of superfields is also equipped with a homological vector field $\delta$ of ghost number $1$ preserving $\Omega$, which we call the \emph{BRST operator}. It is defined as the sum of the commuting vector fields $\hat{D}$ and $\hat{Q}$ obtained from $D$ and $Q$ by acting on superfields $\Phibm\colon T[1]M \to \Ncal$ by the corresponding infinitesimal diffeomorphisms of $T[1]M$ on the left and of $\Ncal$ on the right. The hamiltonian function $S$ (of ghost number zero) of this vector field is then the Batalin-Vilkovisky action functional. More explicitly, this is
\begin{equation}\label{eq:2.12}
S = \int_{T[1]M} \left\{ \tfrac{1}{2} \Phibm^{a} \omega_{ab} \uud \Phibm^{b} + (-1)^{n} \Phibm^* \Theta \right\}  \ud^{n} \theta \, \ud^{n} x.
\end{equation}
If $M$ has no boundary, the condition that the BRST operator $\delta$ be nilpotent is equivalent to what is called the \emph{classical master equation}:
\begin{equation}\label{eq:2.13}
(S,S) = 0.
\end{equation}
Therefore the Batalin-Vilkovisky action $S$ is automatically BRST-invariant, with the BRST transformation rule
\begin{equation}\label{eq:2.14}
\delta \Phibm^{a} = D \Phibm^{a} + Q^{a} (\Phibm),
\end{equation}
where $Q^{a}$ are the components of $Q$ relative to $\{u^{a}\}$. 

The gauge-fixing in the Batalin-Vilkovisky framework corresponds to the choice of a lagrangian submanifold in the space of superfields. Such a choice is typically generated by a ``gauge-fixing fermion'' $\Psi$, which is a functional of the fields $\Phi^{(k)a}$ only and with ghost number $-1$. The lagrangian submanifold is then given by fixing the antifields as 
\begin{equation}
\Phi^{(k)+}_{a} = \frac{\delta \Psi}{\delta \Phi^{(k)a}}.
\end{equation}
To construct a gauge-fixing fermion $\Psi$ one must introduce additional fields. The simplest choice makes use of pairs of fields $\bar{C}^{\alpha}$ and $B^{\alpha}$, for an arbitrary set of indices $\alpha$, with the following set of BRST transformation rules:
\begin{equation}\label{eq:2.16}
\begin{aligned}
\delta \bar{C}^{\alpha} &= B^{\alpha}, \\
\delta B^{\alpha} &= 0. 
\end{aligned}
\end{equation}
Note that here we are implicitly choosing the ghost numbers so that the ghost number of $B^{\alpha}$ is the ghost number of $\bar{C}^{\alpha}$ plus $1$. Along with the fields $\bar{C}^{\alpha}$ and $B^{\alpha}$ we include the corresponding antifields $\bar{C}^{+}_{\alpha}$ and $B^{+}_{\alpha}$. The term to be added to the Batalin-Vilkovisky action $S$ is then
\begin{equation}\label{eq:2.17}
\int_{M} \bar{C}^{+}_{\alpha} B^{\alpha} \: \ud^n x.
\end{equation}
Adding such a term to $S$ does not ruin the classical master equation \eqref{eq:2.13}. For this reason, the fields $\bar{C}^{\alpha}$ and $B^{\alpha}$ form what is called a \emph{trivial pair}. The gauge-fixing is performed by choosing the gauge-fixing fermion
\begin{equation}
\Psi = \int_{M} \bar{C}^{\alpha} F_{\alpha}(\Phi_a^{(k)}) \: \ud^n x,
\end{equation} 
where $F_{\alpha}(\Phi_a^{(k)})$ are a set of gauge-fixing conditions for the  $\Phi_a^{(k)}$. The gauge-fixed action then becomes\footnote{The added term is obtained from \eqref{eq:2.17} by setting $\bar{C}^{+}_{\alpha} = \delta \Psi / \delta \bar{C}^{\alpha}$.}
\begin{equation}
I_{\Psi} = \left( S + \int_{M} \frac{\delta \Psi}{\delta \bar{C}^{\alpha} } B^{\alpha} \: \ud^n x \right)_{\Phi^{(k)+}_{a} = \delta \Psi / \delta \Phi^{(k)a}} .
\end{equation}
One can check that this action is invariant under the BRST transformations, which are given by \eqref{eq:2.14} at $\Phi^{(k)+}_{a} = \delta \Psi/\delta \Phi^{(k)a}$, supplemented with \eqref{eq:2.16}.

We close this subsection with a brief discussion of observables. One of the advantages of the AKSZ construction is that the analysis of classical observables is relatively straightforward. The homological vector field $Q$ on $\mathcal{N}$ defines a structure of a cochain complex on the space of functions on $\Ncal$, whose cohomology we will denote as $H_Q^{\sbullet}(\Ncal)$. Consider a function $f$ on $\mathcal{N}$, and expand $\Phibm^* f$ using the local coordinates of $T[1]M$ as
\begin{equation} \label{eq:2.20}
\Phibm^* f = \sum_{k=0}^{n} W^{(k)}(f),
\end{equation}
where, in the same way as before, we have written $W^{(k)}(f)$ to stand for $\tfrac{1}{k!} W^{(k)}_{i_1 \cdots i_k}(f)\theta^{i_1} \cdots \theta^{i_k}$. From the definition of the BRST operator $\delta$, one finds that its action on $\Phibm^* f$ takes the form
\begin{equation} \label{eq:2.21}
\delta (\Phibm^* f) = D \Phibm^* f + \Phibm^* Q f.
\end{equation}
Thus, if $Qf = 0$, combining equations \eqref{eq:2.20} and \eqref{eq:2.21} results in the relations 
\begin{equation}\label{eq:2.22}
\delta W^{(k)}(f) = D W^{(k-1)}(f),
\end{equation}
valid for $k = 0, 1, \dots, n$. The upshot of this relation is that if $\mu_k$ is a $D$-invariant linear functional on the functions on $T[1]M$, then $\mu_k(W^{(k)}(f))$ is $\delta$-closed and can be regarded as a classical observable. In this way, the cohomology $H_Q^{\sbullet}(\Ncal)$ naturally provides a space of classical observables in the theory.


\section{Topological lagrangian for $\Spin(7)$-instantons}\label{sec:3}
The purpose of this section is to construct a topological quantum field theory capturing the geometry of the moduli space of $\Spin(7)$-instantons. This will be carried out using the Mathai-Quillen formalism in the spirit of Atiyah and Jeffrey. The resulting theory turns out to coincide with the higher dimensional analogues of Donaldson-Witten theory discussed in \cite{A-O-S1997} and \cite{B-K-S1998}. 

\subsection{Moduli space of $\Spin(7)$-instantons}\label{sec:3.1}
As a preparation for our discussion, we will first provide some information on moduli spaces of $\Spin(7)$-instantons. For further details the reader can refer to \cite{Munoz-Shahbazi2020}. 

Let $M$ be an $8$-dimensional manifold with an $\Spin(7)$-structure $\Omega$ and let $P$ be a principal bundle over $M$ whose structure group is a compact semi-simple Lie group $G$ with Lie algebra $\gfrak$. We denote by $\Omega^{k}(M,\ad(P))$ the space of $k$-forms on $M$ with values in the adjoint bundle $\ad(P)= P \times_{\ad} \gfrak$. The spaces $\Omega^{k}(M,\ad(P))$ have a natural $L^2$ inner products defined by the metric on $M$ and an $\ad$-invariant inner product $\tr$ on $\gfrak$, namely 
\begin{equation}\label{eq:3.1} 
\langle \alpha,\beta \rangle_k = \int_M  \tr(\alpha \wedge \ast \beta).
\end{equation}
\sloppy A connection $\omega$ on $P$ naturally induces a covariant exterior derivative $\ud^{\omega} \colon \Omega^{k}(M,\ad(P)) \to \Omega^{k+1}(M,\ad(P))$ and its formal adjoint $\delta^{\omega} \colon \Omega^{k+1}(M,\ad(P)) \to \Omega^{k}(M,\ad(P))$. The curvature of $\omega$ is the element $F_{\omega} \in \Omega^2(M,\ad(P))$ representing the composition $\ud^{\omega} \circ \ud^{\omega} \colon \Omega^{0}(M,\ad(P)) \to  \Omega^{2}(M,\ad(P))$. It satisfies the Bianchi identity $\ud^{\omega}F_{\omega}=0$. 

In \S\ref{sec:2.1}, we have already remarked that there is a decomposition $\Omega^2(M) = \Omega^2_{7}(M) \oplus \Omega^2_{21}(M)$ and projections $\pi_7 \colon \Omega^2(M) \to \Omega^2_{7}(M)$ and $\pi_{21} \colon \Omega^2(M) \to \Omega^2_{21}(M)$. Tensoring with $\ad(P)$ we obtain the orthogonal splitting $\Omega^2(M,\ad(P)) = \Omega^2_{7}(M,\ad(P)) \oplus \Omega^2_{21}(M,\ad(P))$ whose projections we keep on denoting by $\pi_{7}$ and $\pi_{21}$. A connection $\omega$ on $P$ is called a $\Spin(7)$-\emph{instanton} if its curvature $F_{\omega}$ satisfies $\pi_7(F_{\omega})=0$, or, what amounts to the same thing, 
\begin{equation}\label{eq:3.2}
\ast F_{\omega} = - F_{\omega} \wedge \Omega.
\end{equation}
This is a first order equation on $\omega$. It implies a second order equation on $\omega$, upon acting with $\ud^{\omega}$ and using the Bianchi identity to obtain
\begin{equation}
\delta^{\omega} F_{\omega} = - {\ast} (F_{\omega} \wedge W),
\end{equation}
where $W= \ud \Omega$ is the torsion of the $\Spin(7)$-structure. For construction methods and concrete examples of $\Spin(7)$-instantons we refer to \cite{Lewis1999} and \cite{Tanaka2012}. 

We proceed now to review the construction of the moduli space of solutions to \eqref{eq:3.2}. To begin, let us recall that the group of gauge transformations $\Gcal$ might be equivalently understood as the group of all automorphisms of $P$ which cover the identity map over $M$ or as sections of the associated non-linear adjoint bundle $\Ad(P) = P \times_{\Ad} G$. Endowed with the Whitney topology this is an infinite-dimensional Lie group with Lie algebra $\Omega^0(M,\ad(P))$. Let us denote the space of all connections on $P$ by $\Acal$. It is an affine space modelled on $\Omega^1(M,\ad(P))$, and thus its tangent space $T_{\omega} \Acal$ at  each point $\omega \in \Acal$ is naturally identified with $\Omega^1(M,\ad(P))$. The group $\Gcal$ acts on the right on $\Acal$ by the rule $\omega \cdot f = f^* \omega$ for any $\omega \in \Acal$ and any $f \in \Gcal$. It can be shown that this action is smooth and, if $\omega \in \Acal$ is fixed, its derivative in the $\Gcal$ variable can be identified with $\ud^{\omega} \colon \Omega^{0}(M,\ad(P)) \to \Omega^{1}(M,\ad(P))$. Consequently, at each point $\omega \in \Acal$, $T_{\omega} \Acal$ can be split into a ``vertical part'' $\im \ud^{\omega}$, tangent to the orbit of $\Gcal$ through $\omega$, and a ``horizontal part'' $\ker  \delta^{\omega}$, the orthogonal complement of $\im \ud^{\omega}$ with respect to the $L^2$ inner product \eqref{eq:3.1}. On the other hand we can define the isotropy group $\Gamma_{\omega}$ of $\omega$ in $\Gcal$. It is a compact Lie group which is identified with a subgroup of $G$. In fact, $\Gamma_{\omega}$ is the centraliser of the holonomy subgroup of the connection $\omega$ and always contains the centre $Z(G)$ of $G$. Connections $\omega \in \Acal$ with $\Gamma_{\omega}= Z(G)$ are called \emph{irreducible}. We write $\Acal^*$ for the open subset of $\Acal$ consisting of irreducible connections on $P$ and consider the orbit space $\Bcal^*$ arising from the action of $\Gcal$ on $\Acal^*$. Since the centre of $\Gcal$ coincides with $Z(G)$, we should take the quotient group $\Gcal^* = \Gcal/Z(G)$ instead of $\Gcal$ in order to make the action free. Working with appropriate Sobolev spaces of connections (we will not indicate this explicitly in the following) it can be shown that the space $\Bcal^*$ is a smooth Hausdorff Hilbert manifold, and that the natural projection from $\Acal^*$ to $\Bcal^*$ defines a principal bundle with structure group $\Gcal^*$. 

Next define $\Scal^*$ to be the subset of $\Acal^*$ given by all irreducible solutions to \eqref{eq:3.2}. The image of $\Scal^*$ in the orbit space $\Bcal^*$ is called the moduli space of $\Spin(7)$-instantons and will be denoted by $\Mcal$. To each $\omega \in \Scal^*$ we associate the ``fundamental elliptic complex''
\begin{equation}\label{eq:3.4}
\begin{tikzpicture}[scale=1.0,baseline=-0.1cm, inner sep=1mm,>=stealth]
\node (0) at (0,0) {$0$};
\node (1) at (2.5,0)  {$\Omega^{0}(M,\ad(P))$};
\node (2) at (6.2,0)  {$\Omega^{1}(M,\ad(P))$};
\node (3) at (9.9,0)  {$\Omega^{2}_{7}(M,\ad(P))$};
\node (4) at (12.4,0) {$0,$};
\tikzset{every node/.style={fill=white}} 
\draw[->] (0) to node[midway,above,fill=none] {} (1); 
\draw[->] (1) to node[midway,above,fill=none] {\scriptsize$\ud^{\omega}$} (2); 
\draw[->] (2) to node[midway,above,fill=none] {\scriptsize$\pi_7 \circ \ud^{\omega}$} (3);
\draw[->] (3) to node[midway,above,fill=none] {} (4);
\end{tikzpicture}
\end{equation}
which gives information on the infinitesimal behaviour of $\Mcal$ near the equivalence class $[\omega]$ containing $\omega$. Indeed, this complex has finite-dimensional cohomology groups $H^{0}(\omega)$, $H^{1}(\omega)$ and $H^{2}(\omega)$. From the irreducibility of $\omega$, it follows immediately that $H^{0}(\omega)$ is trivial. Thus the principal stratum of $\Mcal$ consists of the equivalence classes $[\omega]$ of those $\omega$ for which $H^{2}(\omega)$ vanishes. For such $\omega$, the tangent space $T_{[\omega]}\Mcal$ can be identified with $H^{1}(\omega)$. 
In particular, by applying the Atiyah-Singer index theorem to the complex \eqref{eq:3.4}, one concludes that, near $[\omega]$, the moduli space $\Mcal$ is a smooth manifold of dimension
\begin{equation}
\dim \Mcal = -(\dim \gfrak) \hat{A}_2(M) + \tfrac{1}{24} \int_M p_1(M) p_1(\ad(P)) - \tfrac{1}{12} \int_M \left(p_1(\ad(P))^2 - 2 p_2(\ad(P))\right). 
\end{equation}
Here $\hat{A}_2(M)$ is the second term of the $\hat{A}$-genus of $M$, $p_1(M)$ is the first Pontryagin class of $M$, and $p_1(\ad(P))$ and $p_2(\ad(P))$ are the first and second Pontryagin class of the adjoint bundle $\ad(P)$, respectively.  


\subsection{Topological quantum field theory realisation}\label{sec:3.2}
In this subsection we will construct a topological quantum field theory which corresponds to the moduli problem of $\Spin(7)$-instantons. As we stressed above, this is accomplished by specifying its topological lagrangian from the Atiyah-Jeffrey realisation of the Mathai-Quillen formalism. All the notation and conventions of \S\ref{sec:3.1} remain in force.

To implement the Mathai-Quillen construction, we need first of all to be able to interpret the moduli space of $\Spin(7)$-instantons $\Mcal$ as the zero locus of a section of a vector bundle. For this purpose, we invoke the fact that the space of irreducible connections $\Acal^*$ can be regarded as the total space of a principal bundle with structure group $\Gcal^*$ over the Hilbert manifold $\Bcal^* = \Acal^* / \Gcal^*$. Since $\Gcal^*$ acts on $\Omega_7^2(M,\ad(P))$, we can form the associated infinite-dimensional vector bundle $\Acal^* \times_{\Gcal^*} \Omega_7^2(M,\ad(P))$. This bundle has a natural section determined by the $\Gcal^*$-equivariant map $S \colon \Acal^* \to \Omega_7^2(M,\ad(P))$ defined by $S(\omega)= \pi_7 (F_{\omega})$. Its zero locus is indeed precisely the moduli space $\Mcal$. For the remainder of this subsection, we assume that we are in the situation in which the dimension of $\Mcal$ is zero. 

We will now work out the appropriate analogue for each of the terms in the exponent of the Atiyah-Jeffrey formula \eqref{eq:2.8}. The first, $-\lVert S(\omega) \rVert^2$, is simple enough. As indicated above, our $S$ is the $\pi_7$-projection of the curvature map $S \colon \Acal^* \to \Omega^{2}_{7}(M,\ad(P))$ and the norm is computed from the $L^2$ inner product $\langle , \rangle_2$ on $\Omega^{2}(M,\ad(P))$ introduced in \eqref{eq:3.1}. Thus,
\begin{equation}\label{eq:3.6}
-\lVert S(\omega) \rVert^2 = - \lVert  \pi_7(F_{\omega}) \rVert^2 = - \int_{M} \tr \left( \pi_7(F_{\omega}) \wedge \ast \pi_7(F_{\omega}) \right). 
\end{equation}
We may also express this another way. We first note that, by virtue of the orthogonality of the decomposition $F_{\omega} = \pi_7(F_{\omega}) + \pi_{21}(F_{\omega})$, we have
\begin{equation}\label{eq:3.7}
 \lVert  F_{\omega} \rVert^2 = \lVert  \pi_7(F_{\omega}) \rVert^2 +  \lVert  \pi_{21}(F_{\omega}) \rVert^2. 
\end{equation}
On the other hand, making use of \eqref{eq:2.4}, one deduces that
\begin{equation}
\begin{aligned}
\tr \left( F_{\omega} \wedge F_{\omega} \wedge \Omega \right) &= 3 \tr \left(\pi_7(F_{\omega}) \wedge \ast \pi_7(F_{\omega}) \right) -  \tr \left(\pi_{21}(F_{\omega}) \wedge \ast \pi_{21}(F_{\omega})\right) \\
&\quad  + 3  \tr \left( \pi_{21}(F_{\omega}) \wedge \ast \pi_7(F_{\omega}) \right) -  \tr \left(\pi_{7}(F_{\omega}) \wedge \ast \pi_{21}(F_{\omega})\right),
\end{aligned}
\end{equation}
and integrating this over $M$, one arrives at
\begin{equation}\label{eq:3.9}
\int_M \tr \left(F_{\omega} \wedge F_{\omega} \wedge \Omega\right) = 3 \lVert  \pi_7(F_{\omega}) \rVert^2 -  \lVert  \pi_{21}(F_{\omega}) \rVert^2.
\end{equation}
Combining \eqref{eq:3.7} and \eqref{eq:3.9} then yields
\begin{equation}\label{eq:3.10}
-\lVert S(\omega) \rVert^2  = - \tfrac{1}{4} \int_{M}\tr \left( F_{\omega} \wedge \ast F_{\omega} \right) - \tfrac{1}{4}\int_M \tr \left(F_{\omega} \wedge F_{\omega} \wedge \Omega\right).
\end{equation}
The first term here is the usual Yang-Mills action, while the second is a topological term. 

To proceed further, it is necessary to ascertain the proper analogues of the maps $C$, $C^*$ and $R$ to the current context. To this end we recall that, at each point $\omega \in \Acal^*$, the map $C_{\omega}$ takes the Lie algebra of $\Gcal^*$, which we have seen may be identified with $\Omega^0(M,\ad (P))$, to the tangent space $T_{\omega} \Acal^*$, which may be identified with $\Omega^1(M,\ad (P))$. An explicit expression for $C_{\omega}$ is obtained by setting
\begin{equation}
C_{\omega} (\xi) =\frac{\ud}{\ud t} \bigg\vert_{t=0} \omega \cdot \exp(-t \xi),
\end{equation}
for each $\xi \in \Omega^0(M,\ad (P))$. Calculating this locally by expanding the exponential, we find that $C_{\omega} = \ud^{\omega} \colon \Omega^0(M,\ad (P)) \to \Omega^1(M,\ad (P))$. From this it follows, in particular, that $C^*_{\omega} = \delta^{\omega} \colon \Omega^1(M,\ad (P)) \to \Omega^0(M,\ad (P))$,
and therefore $R_{\omega} = C^*_{\omega} \circ C_{\omega} = \Delta_0^{\omega} \colon \Omega^0(M,\ad (P)) \to \Omega^0(M,\ad (P))$ is the scalar Laplacian associated to $\omega$.

Armed with this background, we may now consider the term $i (\phi,R_{\omega} \bar{\phi})$. Since here both $\phi$ and $\bar{\phi}$ belong to the Lie algebra of $\Gcal^*$, we must introduce two bosonic fields $\phi,\bar{\phi} \in \Omega^0(M,\ad (P))$ and interpret $(,)$ as the natural $L^2$ inner product $\langle , \rangle_0$ on $\Omega^0(M,\ad (P))$. Hence $i (\phi,R_{\omega} \bar{\phi})$ is expressible in the form
\begin{equation}\label{eq:3.12}
i (\phi,R_{\omega} \bar{\phi}) = i \langle \phi, \Delta_0^{\omega} \bar{\phi} \rangle_0 =  i \int_M \tr \left( \phi  \astDelta_0^{\omega} \bar{\phi} \right).
\end{equation}
Next, let us consider the term $(C_{\omega}^* \psi, \eta)$. Since the map $C_{\omega}^*$ sends $\Omega^1(M,\ad (P))$ to $\Omega^0(M,\ad (P))$, we will need two fermionic fields $\psi \in \Omega^1(M,\ad (P))$ and $\eta \in \Omega^0(M,\ad (P))$ and, as above, interpret $(,)$ as $\langle , \rangle_0$. Thus, we may write
\begin{equation}\label{eq:3.13}
(C_{\omega}^* \psi, \eta) = \langle \delta^{\omega} \psi, \eta \rangle_0 = \langle  \psi, \ud^{\omega}\eta \rangle_1 = \langle \ud^{\omega}\eta, \psi \rangle_1 =  \int_{M} \tr \left( \ud^{\omega}\eta \wedge \ast \psi \right),
\end{equation}
or, if we invoke the identity $\ud \tr \left( \eta {\ast} \psi\right) = \tr \left( \ud^{\omega} \eta \wedge \ast \psi \right) + \tr \left( \eta \ud^{\omega} {\ast} \psi\right)$,
\begin{equation}\label{eq:3.14}
(C_{\omega}^* \psi, \eta) = - \int_{M} \tr \left( \eta \ud^{\omega} {\ast} \psi \right).
\end{equation}
Let us now turn to the term $i \chi^{\mathrm{T}} \ud S_{\omega}(\psi)$. Recall from \S\ref{sec:2.2} that the fermionic variable $\chi$ in the Mathai-Quillen formalism arises from the odd generators of the fibre $V$ of the associated vector bundle. In the present case this fibre is $\Omega^2_{7}(M,\ad(P))$, which implies that we must introduce a fermionic field $\chi \in \Omega^2_{7}(M,\ad(P))$. On the other hand, the differential of the $\pi_7$-projection of the curvature map $S \colon \Acal^* \to \Omega^2_7(M,\ad(P))$ at $\omega \in \Acal^*$ may be identified with the $\pi_7$-projection of the covariant exterior derivative $\pi_7 \circ \ud^{\omega} \colon \Omega^1(M,\ad(P)) \to \Omega^2_{7}(M,\ad(P))$. In particular, this means that $\ud S_{\omega}(\psi) = \pi_7(\ud^{\omega} \psi)$. As for the finite dimensional expression $\chi^{\mathrm{T}} \ud S_{\omega}(\psi)$, we interpret it in terms of the pertinent $L^2$ inner product $\langle , \rangle_2$. With this understanding, $i \chi^{\mathrm{T}} \ud S_{\omega}(\psi)$ may be expressed as
\begin{equation}\label{eq:3.15}
i \chi^{\mathrm{T}} \ud S_{\omega}(\psi) = i \langle \chi , \pi_7(\ud^{\omega} \psi)  \rangle_2 = i \langle   \chi , \ud^{\omega} \psi \rangle_2 =  i \langle   \ud^{\omega} \psi , \chi \rangle_2 =  i \int_M \tr \left( \ud^{\omega} \psi  \wedge \ast \chi   \right).
\end{equation}
We shall now turn to the term $\frac{1}{4} \chi^{\mathrm{T}}(\rho_*\phi) \chi$. From \S\ref{sec:2.2} we recall that the homomorphism $\rho$ gives us the action of $G$ on $V$ that defines the associated vector bundle. In the current situation, $\Gcal^*$, regarded as the space of sections of the associated non-linear adjoint bundle, acts on $\Omega^2_{7}(M,\ad(P))$ pointwise by conjugation. The infinitesimal version of this action is therefore given by $(\rho_*\phi) \chi = [\phi,\chi]$. Hence, in view of our foregoing remarks,
\begin{equation}\label{eq:3.16}
\tfrac{1}{4} \chi^{\mathrm{T}}(\rho_*\phi) \chi = \tfrac{1}{4} \langle \chi, [\phi,\chi] \rangle_2 = \tfrac{1}{4} \int_M \tr \left( \chi \wedge \ast [\phi,\chi]\right) = \tfrac{1}{4} \int_M \tr \left(\phi [\chi,\ast\chi] \right) ,
\end{equation}
where we have used the $\ad$-invariant property of $\tr$ in the last step. 

The only remaining term in \eqref{eq:2.8} is $-i(\ud C_{\omega}^* (\psi,\psi), \bar{\phi})$ and this requires a bit of work. In accordance to the remarks made above, the map $C^*$ is a $1$-form on $\Acal^*$ which takes values in $\Omega^0(M,\ad(P))$. To evaluate the exterior derivative $\ud C$ at $\omega \in \Acal^*$ we proceed as follows. Fix two tangent vectors $\psi_1,\psi_2 \in T_{\omega} \Acal^*$. Since $\Acal$ is an affine space and $\Acal^*$ is open in $\Acal$, we may think of $\psi_1$ and $\psi_2$ as being constant vector fields on $\Acal^*$. Therefore, by a well known formula, we have
\begin{equation}\label{eq:3.17}
\ud C^* (\psi_1,\psi_2) = \psi_1 (C^* \psi_2) - \psi_2 (C^* \psi_1) - C^*([\psi_1,\psi_2]) = \psi_1 (C^* \psi_2) - \psi_2 (C^* \psi_1),
\end{equation}
where here $C^* \psi_i$ denotes the function from $\Acal^*$ to $\Omega^0(M,\ad(P))$ sending $\omega'$ to $C_{\omega'}^* \psi_i$, and where the Lie bracket of $\psi_1$ and $\psi_2$ is zero because they are constant vector fields. Now, let us determine the expression for $\psi_1 (C^* \psi_2)$ at $\omega \in \Acal^*$. On unravelling the definitions we get
\begin{equation}\label{eq:3.18}
\psi_1 (C^* \psi_2)(\omega) = \psi_1 (C_{\omega}^* \psi_2) = \frac{\ud}{\ud t} \bigg\vert_{t=0}  C^*_{\omega + t \psi_1} (\psi_2) = \frac{\ud}{\ud t} \bigg\vert_{t=0} \delta^{\omega + t \psi_1} \psi_2. 
\end{equation}
To calculate $\delta^{\omega + t \psi_1} \psi_2$ it is helpful to consider the map $b_{\psi_1} \colon \Omega^0(M,\ad(P)) \to \Omega^1(M,\ad(P))$ defined by $b_{\psi_1}(\xi) = [\psi_1,\xi]$. In terms of such a map, we may write, for any $\xi \in \Omega^0(M,\ad(P))$,
\begin{equation}
\ud^{\omega + t \psi_1} \xi = \ud^{\omega} \xi + t b_{\psi_1}(\xi).
\end{equation}
It then follows that
\begin{equation}
\delta^{\omega + t \psi_1} \psi_2 = \delta^{\omega} \psi_2 + t b^*_{\psi_1}(\psi_2),
\end{equation}
where $b^*_{\psi_1} \colon \Omega^1(M,\ad(P)) \to \Omega^0(M,\ad(P))$ is the formal adjoint of $b_{\psi_1}$. We claim that
\begin{equation}\label{eq:3.21}
b^*_{\psi_1}(\psi_2) = {\ast} [\psi_1,\ast \psi_2].
\end{equation}  
Indeed, for any $\xi \in \Omega^0(M,\ad(P))$,
\begin{equation}
\langle b_{\psi_1}(\xi),\psi_2 \rangle_1  =  \int_M \tr \left([\psi_1,\xi] \wedge \ast \psi_2 \right) =  \int_M \tr \left(\xi \wedge [\psi_1, \ast\psi_2] \right) = \langle \xi , \ast [\psi_1, \ast\psi_2] \rangle_0,
\end{equation}
from which \eqref{eq:3.21} is readily obtained. Putting everything together in \eqref{eq:3.18}, we obtain
\begin{equation}
\psi_1 (C^* \psi_2)(\omega) = \ast [\psi_1, \ast\psi_2].
\end{equation}
By interchanging $\psi_1$ with $\psi_2$ we similarly find that
\begin{equation}
\psi_2 (C^* \psi_1)(\omega) = \ast [\psi_2, \ast\psi_1] = - {\ast} [\psi_1, \ast\psi_2] .
\end{equation}
Thus the equality \eqref{eq:3.17} implies that 
\begin{equation}
\ud C_{\omega}^* (\psi_1,\psi_2) =2  {\ast} [\psi_1, \ast\psi_2].
\end{equation}
Using this fact, we can finally determine the sought-after term as
\begin{equation}\label{eq:3.26}
-i(\ud C_{\omega}^* (\psi,\psi), \bar{\phi}) = - 2 i \langle  {\ast} [\psi, \ast\psi], \bar{\phi} \rangle_0 =  - 2 i \langle \bar{\phi},  {\ast} [\psi, \ast\psi] \rangle_0 = - 2 i  \int_M \tr \left(  \bar{\phi} [\psi, \ast\psi]  \right).
\end{equation}

We are now in a position of writing out the topological quantum field theory associated to the moduli space of $\Spin(7)$-instantons. In the first place, in accordance with our previous considerations, each choice of the three bosonic fields $\phi$, $\bar{\phi}$, $\omega$ and the three fermonic fields $\psi$, $\eta$, $\chi$ corresponds to a field configuration. In the second place, following the prescription of Atiyah and Jeffrey, the exponent appearing in \eqref{eq:2.8} must be identified as minus the action, $-S$, of the theory. With the aid of equations \eqref{eq:3.10}, \eqref{eq:3.12}, \eqref{eq:3.14}, \eqref{eq:3.15}, \eqref{eq:3.16} and \eqref{eq:3.26}, it is found that
\begin{equation}\label{eq:3.27}
\begin{aligned}
S = \int_M \tr &\left\{ \tfrac{1}{4}  F_{\omega} \wedge {\ast} F_{\omega} + \tfrac{1}{4} F_{\omega} \wedge F_{\omega} \wedge \Omega - i \ud^{\omega} \psi  \wedge \ast \chi - \tfrac{1}{4} \phi [\chi,\ast\chi] \right. \\
&\left. \!\!\phantom{\tfrac{1}{4}} + 2 i  \bar{\phi} [\psi, \ast\psi]  - i  \phi  \astDelta_0^{\omega} \bar{\phi} + \eta \ud^{\omega} {\ast} \psi \right\}.
\end{aligned}
\end{equation}
And in the third place, the integral expression for the Euler number of the infinite-dimensional vector bundle $\Acal^* \times_{\Gcal^*} \Omega^2_7(M,\ad(P))$, defined via the Atiyah-Jeffrey formula \eqref{eq:2.8}, can be interpreted as the partition function $Z$ of the theory. That is,
\begin{equation}
Z = \int \exp(-S) \, \Dcal\!\psi \, \Dcal\!\eta \, \Dcal\!\chi \, \ud \phi \, \ud \bar{\phi}  \, \ud \omega,
\end{equation}
with $\Dcal\!\psi \, \Dcal\!\eta \, \Dcal\!\chi \, \ud \phi \, \ud \bar{\phi}  \, \ud \omega$ here denoting the integration measure over all field configurations. Notice that we have omitted any constants that might correspond to $(2\pi)^{-d}(2\pi)^{-m}$ since both $d$ and $m$ are infinite under the present assumptions. It goes without saying that the quantum field theory so obtained is manifestly topological. 

Before leaving this subsection, it is instructive to compare our action \eqref{eq:3.27} with the actions previously obtained in \cite{A-O-S1997} and \cite{B-K-S1998}. For this, let us introduce a system of local coordinates $x^1,\dots,x^{8}$ on $M$ and use $\mu,\nu, \ldots$ as summation indices running from $1$ to $8$. Induced local trivialisations $\ud x^{\mu}$ then arise on $T^*M$ so that the riemannian metric $g$ can be represented by its components $g_{\mu\nu}$. Coordinate indices are lowered and raised by using $g_{\mu\nu}$ and its inverse $g^{\mu\nu}$, respectively. We write $\lvert g\rvert$ for the determinant of $g$ and $\sqrt{\lvert g\rvert} \ud^8x = \sqrt{\lvert g\rvert} \ud x^{1} \wedge \cdots \wedge \ud x^{8}$ for the distinguished volume form on $M$. We also adopt the shorthand notation $\partial_{\mu}$ for the partial derivative $\frac{\partial}{\partial x^{\mu}}$.

Relative to the given system of local coordinates, the connection $\omega$ is specified by giving a differential $1$-form $A = A_{\mu} \ud x^{\mu}$ that takes values in $\gfrak$. We can then express the curvature in the form $F_{\omega} = \frac{1}{2} F_{\mu \nu} \ud x^{\mu} \wedge \ud x^{\nu}$, in which $F_{\mu\nu} = \partial_{\mu} A_{\nu} - \partial_{\nu} A_{\mu} + [A_{\mu},A_{\nu}]$. Also, the fermionic fields $\chi$ and $\psi$ can be written as $\psi = \psi_{\mu} \ud x^{\mu}$ and $\chi = \frac{1}{2} \chi_{\mu \nu} \ud x^{\mu} \wedge \ud x^{\nu}$, with both $\psi_{\mu}$ and $\chi_{\mu\nu}$ taking values in $\gfrak$. By applying the Hodge star operator, one finds
\begin{equation} 
\begin{aligned}
\ast F_{\omega} &= \tfrac{1}{6!} \sqrt{\lvert g\rvert} F^{\mu\nu} \varepsilon_{\mu\nu \rho_1 \cdots \rho_6} \ud x^{\rho_1} \wedge \cdots \wedge \ud x^{\rho_6}, \\
\ast \psi &= \tfrac{1}{7!} \sqrt{\lvert g\rvert} \psi^{\mu} \varepsilon_{\mu \rho_1 \cdots \rho_7} \ud x^{\rho_1} \wedge \cdots \wedge \ud x^{\rho_7},\\
\ast \chi &= \tfrac{1}{6!} \sqrt{\lvert g\rvert} \chi^{\mu\nu} \varepsilon_{\mu\nu \rho_1 \cdots \rho_6} \ud x^{\rho_1} \wedge \cdots \wedge \ud x^{\rho_6},
\end{aligned}
\end{equation}
where $\varepsilon_{\mu_1 \cdots \mu_8}$ is the antisymmetric Levi-Civita symbol. Furthermore,
\begin{equation}
\begin{aligned}
\ud^{\omega} \psi &= \tfrac{1}{2} (D_{\mu} \psi_{\nu} - D_{\nu} \psi_{\mu}) \ud x^{\mu} \wedge \ud x^{\nu}, \\
\ud^{\omega} {\ast} \psi & = D_{\mu} \psi^{\mu} \sqrt{\lvert g\rvert} \ud^8 x,\\
{\ast} \Delta_0^{\omega} \bar{\phi} &= D_{\mu} D^{\mu} \bar{\phi} \sqrt{\lvert g\rvert} \ud^8 x,
\end{aligned}
\end{equation}
 whereby the covariant derivative is $D_{\mu} = \partial_{\mu} + [A_{\mu},]$. We too, as a matter of convenience, write $\ast(F_{\omega} \wedge \Omega) = \frac{1}{2} \tilde{F}_{\mu\nu} \ud x^{\mu} \wedge \ud x^{\nu}$, in which $\tilde{F}_{\mu\nu} = \frac{1}{2} \Omega_{\mu\nu\rho\sigma} F^{\rho\sigma}$. Combining all these with the expression for $S$ given by \eqref{eq:3.27} we easily obtain
\begin{equation}
\begin{aligned}
S = \int_M \tr &\left\{ \tfrac{1}{8} F_{\mu\nu}F^{\mu\nu} + \tfrac{1}{8} F_{\mu\nu}\tilde{F}^{\mu\nu} - i D_{\mu} \psi_{\nu} \chi^{\mu\nu} - \tfrac{1}{8} \phi [\chi_{\mu\nu},\chi^{\mu\nu}]  \right. \\
&\left. \!\!\phantom{\tfrac{1}{4}} +2i \bar{\phi}[\psi_{\mu},\psi^{\mu}] - i \phi D_{\mu} D^{\mu} \bar{\phi} + \eta D_{\mu} \psi^{\mu}  \right\} \sqrt{\lvert g\rvert} \ud^8 x.
\end{aligned}
\end{equation}
On close inspection it can be seen that this formula encompasses all the terms in the actions found in  \cite{A-O-S1997} and \cite{B-K-S1998}, albeit with different coefficients. To better conform to the latter, one simply needs to rescale the $\ad$-invariant inner product $\tr$ by a factor of $2$ and the fields $\phi$, $\bar{\phi}$ and $\psi$ by a factor of $\frac{1}{2}$.


\subsection{Observables of the topological quantum field theory}\label{sec:3.3}
After having defined the field content and the action of our topological quantum field theory, the next natural step is to find an interesting set of metric independent observables. On general grounds, these observables are classified according to the cohomology of the BRST operator. To incorporate this operator correctly into our discussion, one must resort to the AKSZ formalism, which is guaranteed to produce a BRST invariant action. We will do this in the next section. In the meantime, we take a more prosaic approach.

Similar to the Donaldson-Witten theory \cite{marino2003introduction,stora1997algebraic}, we will employ an infinite-dimensional version of the Cartan model for the equivariant cohomology on the space of fields which gives the BRST symmetry of the theory. What this means, precisely, is that the BRST operator $\delta$ can be realised as the Cartan exterior derivative for the $\Gcal^*$-equivariant cohomology of $\Acal^*$. For completeness, we include in appendix~\ref{app:A} a short review of the standard finite-dimensional construction. The relevant ingredients are as follows. Let $\Lie(\Gcal^*)$ be the Lie algebra of $\Gcal^*$, which, as we already know, is identified with $\Omega^0(M,\ad(P))$. As per the notation in appendix~\ref{app:A}, we write $\Pol^{\sbullet}(\Lie(\Gcal^*))$ for the algebra of polynomial functions on $\Lie(\Gcal^*)$, and $\Omega^{\sbullet}(\Acal^*)$ for the algebra of differential forms on $\Acal^*$. The algebra of $\Gcal^*$-equivariant differential forms, denoted $\Omega^{\sbullet}_{\Gcal^*}(\Acal^*)$, then consists of the $\Gcal^*$-invariant elements of the tensor product $\Pol^{\sbullet}(\Lie(\Gcal^*))\otimes \Omega^{\sbullet}(\Acal^*)$, where $\Gcal^*$ acts on $\Lie(\Gcal^*)$ (and hence on $\Pol^{\sbullet}(\Lie(\Gcal^*))$) via the adjoint action and naturally on $\Omega^{\sbullet}(\Acal^*)$; it carries a grading where we ``double the degrees’’ in $\Pol^{\sbullet}(\Lie(\Gcal^*))$. Thus defined, the Cartan exterior derivative, here denoted $\delta$ in accordance with its identification with the BRST operator, is obtained by adapting formula~\eqref{eq:A.2} to the infinite-dimensional setting at hand, so that for any $\alpha \in \Pol^{\sbullet}(\Lie(\Gcal^*)) \otimes  \Omega^{\sbullet}(\Acal^*)$ and $\phi \in \Lie(\Gcal^*)$ one has
\begin{equation}\label{eq:3.32}
(\delta \alpha)(\phi) = \ud (\alpha(\phi)) - i_{\phi^{\#}}(\alpha(\phi)),
\end{equation}
where $\phi^{\#}$ is the vector field on $\Acal^*$ adapted from equation~\eqref{eq:A.3} (with a sign change since here we work with a right action rather than the left action), resulting in
\begin{equation}\label{eq:3.33}
\phi^{\#}(\omega) = \frac{\ud}{\ud t} \bigg\vert_{t=0} \omega \cdot \exp(t \phi) = - \ud^{\omega} \phi,
\end{equation}
and $i_{\phi^{\#}}$ refers to interior product by $\phi^{\#}$. It can be verified that $\delta $ increases the degree by one and squares to zero on $\Omega^{\sbullet}_{\Gcal^*}(\Acal^*)$. The $\Gcal^*$-equivariant cohomology of $\Acal^*$ is then defined to be the cohomology of $\delta$ acting on $\Omega^{\sbullet}_{\Gcal^*}(\Acal^*)$.

Let us now turn to the observables of the theory. A natural set of topological observables is based on forms on $M$, which can be grouped into families labeled by a positive integer $n$. These forms can be obtained in the following way. Consider the trivial extension of the action of $\Gcal^*$ on $\Acal^*$ to the product $M \times \Acal^*$, and write $\Omega^{\sbullet}_{\Gcal^*}(M \times \Acal^*)$ for the associated algebra of $\Gcal^*$-equivariant differential forms. Clearly, this algebra can be thought of as the tensor product of the algebra of differential forms on $M$ with the algebra of $\Gcal^*$-equivariant forms on $\Acal^*$. Thus, it carries a natural bigrading, a $(p,q)$-form referring to a $p$-form on $M$ and a $\Gcal^*$-equivariant $q$-form on $\Acal^*$. It is also worth noting that the usual exterior differential $\ud$ and the BRST operator $\delta$ are graded derivations with bigradings $(1,0)$ and $(0,1)$, respectively. With this in mind let us define $W_2^{(0)} \in \Omega^{0,4}_{\Gcal^*}(M \times \Acal^*)$ by putting
\begin{equation}\label{eq:3.34}
W_2^{(0)}(\phi) = \tfrac{1}{2}  \tr \phi^2,
\end{equation} 
for any $\phi \in \Lie(\Gcal^*)$. We claim that $\ud W_2^{(0)}$ is $\delta$-exact. To justify the claim, let $W_2^{(1)} \in \Omega^{1,3}_{\Gcal^*}(M \times \Acal^*)$ be defined by
\begin{equation}\label{eq:3.35}
(W_2^{(1)}(\phi))_{\omega}(\psi) =  \tr \left( \phi \wedge \psi \right),
\end{equation} 
for any $\phi \in \Lie(\Gcal^*)$, $\omega \in \Acal^*$ and $\psi \in T_{\omega} \Acal^*$. We want to calculate $\delta W_2^{(1)}$ at a fixed $\phi \in \Lie(\Gcal^*)$. According to \eqref{eq:3.32}, this is
\begin{equation}\label{eq:3.36}
(\delta W_2^{(1)})(\phi) = \ud (W_2^{(1)}(\phi)) - i_{\phi^{\#}} (W_2^{(1)}(\phi)).
\end{equation}
Now, a look at the definition of \eqref{eq:3.35} shows that $W_2^{(1)}(\phi)$ is a constant $1$-form on $\Acal^*$. Consequently, $\ud (W_2^{(1)}(\phi)) = 0$. On the other hand, for any $\omega \in \Acal^*$, \eqref{eq:3.33} implies that
\begin{equation}
(i_{\phi^{\#}} (W_2^{(1)}(\phi)))(\omega)  = - (W_2^{(1)}(\phi))_{\omega}(\ud^{\omega}\phi)  
=  - \tr \left( \phi \wedge \ud^{\omega}\phi \right) = - \tfrac{1}{2} \tr \left(  \ud^{\omega}\phi^2 \right) = - \ud \left( \tfrac{1}{2} \tr \phi^2 \right)
\end{equation}
or, using \eqref{eq:3.34},
\begin{equation}\label{eq:3.38}
(i_{\phi^{\#}} (W_2^{(1)}(\phi)))(\omega) = -\ud (W_2^{(0)}(\phi)).
\end{equation}
From \eqref{eq:3.36} and \eqref{eq:3.38} it follows then that
\begin{equation}
\ud W_2^{(0)} = \delta W_2^{(1)},
\end{equation}
and hence the claim holds true. This is the first relation which we have been seeking. To continue with the construction, we now must check that $\ud W_2^{(1)}$ is also $\delta$-exact. For this purpose, consider the element $W_2^{(2)} \in  \Omega^{2,2}_{\Gcal^*}(M \times \Acal^*)$ which as a differential form on $\Acal^*$ decomposes as $W_2^{(2)} = W_2^{(2,0)} + W_2^{(2,2)}$ with homogeneous components defined by means of
\begin{align}
(W_2^{(2,0)}(\phi))(\omega) &=  \tr \left( \phi \wedge F_{\omega} \right), \label{eq:3.40} \\
(W_2^{(2,2)}(\phi))_{\omega}(\psi_1,\psi_2) &=   \tr \left( \psi_1 \wedge \psi_2 \right), \label{eq:3.41} 
\end{align}
for any $\phi \in \Lie(\Gcal^*)$, $\omega \in \Acal^*$ and $\psi_1,\psi_2 \in T_{\omega} \Acal^*$. Again, at a fixed $\phi \in \Lie(\Gcal^*)$, the formula \eqref{eq:3.32} gives
\begin{equation}\label{eq:3.42}
(\delta W_2^{(2)})(\phi) = \ud (W_2^{(2,0)}(\phi)) + \ud (W_2^{(2,2)}(\phi)) - i_{\phi^{\#}} (W_2^{(2,0)}(\phi)) - i_{\phi^{\#}} (W_2^{(2,2)}(\phi)).
\end{equation}
We note that $\ud (W_2^{(2,2)}(\phi)) = 0$ because from the definition \eqref{eq:3.41} it is clear that $W_2^{(2,2)}(\phi)$ is a constant $2$-form on $\Acal^*$. Moreover, it is evident that $ i_{\phi^{\#}} (W_2^{(2,0)}(\phi)) = 0$ since $W_2^{(2,0)}(\phi)$ is a $0$-form on $\Acal^*$. We are thus left to determine $\ud (W_2^{(2,0)}(\phi))$ and $i_{\phi^{\#}} (W_2^{(2,2)}(\phi))$. For any $\omega \in \Acal^*$ and $\psi \in T_{\omega} \Acal^*$, we find that
\begin{equation}
\begin{aligned}
(\ud (W_2^{(2,0)}(\phi)))_{\omega}(\psi) &=  \frac{\ud}{\ud t} \bigg\vert_{t=0} (W_2^{(2,0)}(\phi))(\omega + t \psi)  \\
&= \frac{\ud}{\ud t} \bigg\vert_{t=0}  \tr \left( \phi \wedge F_{\omega + t \psi} \right) \\
& =  \tr \left( \phi \wedge \ud^{\omega} \psi \right),
\end{aligned}
\end{equation}
and 
\begin{equation}
\begin{aligned}
(i_{\phi^{\#}} (W_2^{(2,2)}(\phi)))_{\omega}(\psi) &= -(W_2^{(2,2)}(\phi))_{\omega}(\ud^{\omega}\phi,\psi) \\
& = -   \tr \left( \ud^{\omega}\phi  \wedge \psi \right),
\end{aligned}
\end{equation}
where in the last equality of the first equation we have used the fact that $\frac{\ud}{\ud t} \big\vert_{t=0} F_{\omega + t \psi} = \ud^{\omega} \psi$. Combining these two results with \eqref{eq:3.42}, we see that
\begin{equation}
((\delta W_2^{(2)})(\phi))_{\omega}(\psi) = \tr \left( \phi \wedge \ud^{\omega} \psi \right) +  \tr \left( \ud^{\omega}\phi  \wedge \psi \right)  = \ud   \tr \left( \phi  \wedge \psi \right) .
\end{equation}
Recalling \eqref{eq:3.35}, the above equality yields
\begin{equation}
 \ud W_2^{(1)} = \delta W_2^{(2)},
\end{equation}
as was to be shown. At this point we can already anticipate that the next step will be to verify the requirement that $\ud W_2^{(2)}$ is $\delta$-exact. To this end we consider the element $W_2^{(3)}  \in  \Omega^{3,1}_{\Gcal^*}(M \times \Acal^*)$ which is given by
\begin{equation} \label{eq:3.47}
(W_2^{(3)}(\phi))_{\omega}(\psi) =  \tr \left( \psi \wedge F_{\omega} \right),
\end{equation}
for any $\phi \in \Lie(\Gcal^*)$, $\omega \in \Acal^*$ and $\psi \in T_{\omega} \Acal^*$. Then \eqref{eq:3.32} tells us that, at a fixed $\phi \in \Lie(\Gcal^*)$, 
\begin{equation}\label{eq:3.48}
(\delta W_2^{(3)})(\phi) =  \ud (W_2^{(3)}(\phi))  - i_{\phi^{\#}} (W_2^{(3)}(\phi)).
\end{equation}
Furthermore, for any $\omega \in \Acal^*$, it is found that
\begin{equation}\label{eq:3.49}
(i_{\phi^{\#}} (W_2^{(3)}(\phi)))(\omega) = - \ud  \tr \left( \phi \wedge F_{\omega} \right).
\end{equation}
Additionally, for any $\omega \in \Acal^*$ and $\psi_1,\psi_2 \in T_{\omega}\Acal^*$, if we think of $\psi_1$ and $\psi_2$ as constant vector fields on $\Acal^*$, then
\begin{equation}\label{eq:3.50}
\ud (W_2^{(3)}(\phi))_{\omega}(\psi_1,\psi_2) = \psi_1 (W_2^{(3)}(\phi) \psi_2) - \psi_2 (W_2^{(3)}(\phi) \psi_1),
\end{equation}
where, as before, $W_2^{(3)}(\phi) \psi_i$ is the function from $\Acal^*$ to $\Omega^{3}(M)$ taking $\omega$ to $(W_2^{(3)}(\phi))_{\omega}(\psi_i)$ as defined using \eqref{eq:3.47}. Proceeding with a direct calculation,
\begin{equation}\label{eq:3.51}
\begin{aligned}
\psi_1 (W_2^{(3)}(\phi) \psi_2)(\omega) &= \frac{\ud}{\ud t} \bigg\vert_{t=0} (W_2^{(3)}(\phi))_{\omega + t \psi_1}(\psi_2)  \\
&= \frac{\ud}{\ud t} \bigg\vert_{t=0}  \tr \left( \psi_2 \wedge F_{\omega + t \psi_1} \right) \\
&=  \tr \left( \psi_2 \wedge \ud^{\omega} \psi_1 \right),
\end{aligned}
\end{equation}
and, just interchanging $\psi_1$ and $\psi_2$,
\begin{equation}
\psi_2 (W_2^{(3)}(\phi) \psi_1)(\omega)  =  \tr \left( \psi_1 \wedge \ud^{\omega} \psi_2 \right).
\end{equation}
With the aid of \eqref{eq:3.50} we thus obtain
\begin{equation}
\ud (W_2^{(3)}(\phi))_{\omega}(\psi_1,\psi_2) =  \tr \left(  \psi_2 \wedge \ud^{\omega} \psi_1 \right) -  \tr \left( \psi_1 \wedge \ud^{\omega} \psi_2 \right) = \ud \tr \left( \psi_1 \wedge  \psi_2 \right).
\end{equation}
This, together with \eqref{eq:3.47}, in combination with \eqref{eq:3.48}, reveals a decomposition $\delta W_2^{(3)} = \delta W_2^{(3,0)} + \delta W_2^{(3,2)} \in  \Omega^{3,2}_{\Gcal^*}(M \times \Acal^*)$, with its homogeneous components being specified by
\begin{equation}
((\delta W_2^{(3,0)})(\phi))(\omega) = - (i_{\phi^{\#}} (W_2^{(3)}(\phi)))(\omega) = \ud  \tr \left( \phi \wedge F_{\omega} \right),  
\end{equation}
 and 
 \begin{equation}
 ((\delta W_2^{(3,2)})(\phi))_{\omega}(\psi_1,\psi_2) = \ud (W_2^{(3)}(\phi))_{\omega}(\psi_1,\psi_2) = \ud  \tr \left(  \psi_1 \wedge \psi_2 \right).
 \end{equation}
 Bringing to mind \eqref{eq:3.40} and \eqref{eq:3.41}, the previous equations lead to
\begin{equation}
\ud W_2^{(2)} = \delta W_2^{(3)},
\end{equation}
as desired. Now, although it may seem repetitive, we are on the verge of our final step, which involves verifying that $\ud W_2^{(3)}$ is $\delta$-exact. In this regard, we consider the element $W_2^{(4)}  \in \Omega^{4,0}_{\Gcal^*}(M \times \Acal^*)$ defined as
\begin{equation}
(W_2^{(4)}(\phi))(\omega) = \tfrac{1}{2} \tr \left(  F_{\omega} \wedge F_{\omega} \right),
\end{equation}
for any $\phi \in \Lie(\Gcal^*)$ and $\omega \in \Acal^*$. We want to calculate $\delta W_2^{(4)}$ at a fixed $\phi \in \Lie(\Gcal^*)$. By use of \eqref{eq:3.32} we see that
\begin{equation}\label{eq:3.58}
(\delta W_2^{(4)})(\phi) = \ud (W_2^{(4)}(\phi)),
\end{equation}
where, for the same reason as before, we gather that $i_{\phi^{\#}} (W_2^{(4)}(\phi)) = 0$. Also, by following the same line of reasoning as previously, for any $\omega \in \Acal^*$ and $\psi \in T_{\omega} \Acal^*$,
\begin{equation}
 \ud (W_2^{(4)}(\phi))_{\omega}(\psi) =   \tr \left(  \ud^{\omega} \psi \wedge F_{\omega} \right) = \ud \tr \left( \psi \wedge F_{\omega} \right) .
\end{equation}
When this, together with \eqref{eq:3.58}, is combined with \eqref{eq:3.47}, it leads directly to
\begin{equation}
\ud W_2^{(3)} = \delta W_2^{(4)},
\end{equation}
in agreement with our purpose. Finally, invoking the Bianchi identity, we observe that $\ud W_2^{(4)} = 0$, and thus the process comes to a natural end.

In summary, we have produced quantities $W_2^{(k)} \in \Omega^{k,4-k}_{\Gcal^*}(M \times \Acal^*)$ for $k = 0,1,2,3,4$, which satisfy the relations
\begin{equation}\label{eq:3.61}
\delta W_2^{(k)} = \ud W_2^{(k-1)}.
\end{equation}
By picking a $k$-homology cycle $\gamma$ on $M$ we can then construct an element $\Ocal_2^{(k)}(\gamma) \in \Omega^{4-k}_{\Gcal^*}(\Acal^*)$ via the rule
\begin{equation}
\Ocal_2^{(k)}(\gamma) = \int_{\gamma} W_2^{(k)}.
\end{equation}
On use of \eqref{eq:3.61} it follows that
\begin{equation}
\delta \Ocal_2^{(k)}(\gamma) = \int_{\gamma} \delta W_2^{(k)} = \int_{\gamma} \ud W_2^{(k-1)} = \int_{\partial \gamma}  W_2^{(k-1)} = 0, 
\end{equation}
showing that $\Ocal_2^{(k)}(\gamma)$ is $\delta$-closed and therefore an observable. In addition, $\Ocal_2^{(k)}(\gamma)$ depends only on the homology class of $\gamma$. For if $\gamma$ is a boundary, say $\gamma = \partial \beta$, then 
\begin{equation}
\Ocal_2^{(k)}(\gamma) = \int_{\partial \beta} W_2^{(k)} = \int_{\beta} \ud W_2^{(k)} = \int_{\beta} \delta W_2^{(k+1)} = \delta \int_{\beta} W_2^{(k+1)}.
\end{equation}
Hence, if $\gamma$ is trivial in homology, we must have that $\Ocal_2^{(k)}(\gamma)$ is $\delta$-exact. We have thus found an assignment of $k$-homology classes of $M$ to BRST equivalence classes of observables. In the next section, we shall see that the ensuing relations in the hierarchy of equations \eqref{eq:3.61} which generate these observables are not merely fortuitous. Instead, they stem from a more fundamental property of the AKSZ construction.

In the situation where the $\Spin(7)$-structure $\Omega$ is assumed to be integrable, it was proposed in \cite{A-O-S1997} and \cite{B-K-S1998} that, in addition to the above quantities, one can consider a further four quantities which can then be used to define observables. These are the elements $W_2^{(k)} \in \Omega^{k,8-k}_{\Gcal^*}(M \times \Acal^*)$, with $k$ running over the values $5,6,7,8$, determined by the equations
\begin{equation}
W_2^{(k)} = W_2^{(k-4)} \wedge \Omega.
\end{equation} 
They are obtainable from equations similar to those above by replacing $W_2^{(0)}$ with $W_2^{(0)} \wedge \Omega$, since the $\Spin(7)$-structure $\Omega$ is both $\ud$-closed and $\delta$-closed. The quantities $W_2^{(k)}$ so defined also satisfy the relations \eqref{eq:3.61}, and as a consequence, one can construct a set of observables $\Ocal_2^{(k)}(\gamma) \in \Omega^{8-k}_{\Gcal^*}(\Acal^*)$ by integrating each of them over an appropriate $k$-homology cycle $\gamma$ on $M$.


\section{AKSZ formulation}\label{sec:4}
In this section we are concerned with the AKSZ formulation of the topological quantum field theory for $\Spin(7)$-instantons presented in the preceding section. To be more specific, we shall make use of the AKSZ prescription to generate a Batalin-Vilkovisky action for an $8$-dimensional field theory which upon gauge fixing coincides with the action \eqref{eq:3.27}. 

\subsection{Batalin-Vilkovisky action}\label{sec:4.1}
We will commence our discussion by considering the derivation of the Batalin-Vilkovisky action, which applies to our case of interest. To this end, we simply follow the steps outlined in \S\ref{sec:2.3}.

Suppose that we are given the data of \S\ref{sec:3.1}, namely, an $8$-dimensional manifold $M$ with a $\Spin(7)$-structure $\Omega$ and a principal bundle $P$ over $M$ whose structure group is a compact semi-simple Lie group $G$ with Lie algebra $\gfrak$. We again equip $\gfrak$ with an $\ad$-invariant inner product $\tr$. The first aspect to consider is the choice of a source manifold, specified here as the shifted tangent bundle $T[1]M$. As indicated in \S\ref{sec:2.3}, on this graded manifold, we have a homological vector field $D$ of degree $1$ and a non-degenerate $D$-invariant measure $\mu$ of degree $-8$. By choosing local coordinates ${x^{\mu}}$ on $M$ along with their odd counterparts ${\theta^{\mu}}$, we can write $D = \theta^{\mu} \partial_{\mu}$ and $\mu = \ud^8 \theta \, \ud^8 x$. The second aspect entails the target manifold, and to properly identify it, it is useful to present some (hopefully) clarifying remarks at the outset.

Let us denote by $\tau_{[1]}\colon T[1]M \to M$ the bundle projection of $T[1]M$ onto $M$. Then, we can consider the pullback bundle $\tau_{[1]}^* P$ of $P$ by $\tau_{[1]}$, which can be given the structure of a principal bundle over $T[1]M$ with structure group $G$. Consequently, just as in the ordinary case, we can form the associated adjoint bundle $\ad(\tau_{[1]}^* P) = \tau_{[1]}^* P \times_{\ad} \gfrak$. Notice that the degree shifting of this bundle by $n$ is achieved by replacing $\gfrak$ with $\gfrak[n]$. Following \cite{zucchini2008gauging}, it is also feasible to define the so-called bundle of generalised connections on $\tau_{[1]}^* P$ over $T[1]M$, denoted here by $\Ccal(\tau_{[1]}^* P)$. This, as the name indicates, is the natural bundle whose global sections are identified with generalised connections on $\tau_{[1]}^* P$. Analogously to the familiar construction, it proves to be an affine bundle over $T[1]M$ modelled over the vector bundle $T^*[1]T[1]M \otimes \ad(\tau_{[1]}^* P)[1]$.

Next consider the graded manifold $T[4](\gfrak[1] \oplus \gfrak[2]) = (\gfrak[1] \oplus \gfrak[2]) \oplus (\gfrak[5] \oplus \gfrak[6])$. This is a graded symplectic manifold of degree $7$ with canonical symplectic form $\omega$ that, using Lie algebra coordinates ${c,\phi}$ of degrees $1$ and $2$ together with their odd counterparts ${\rho,\sigma}$ of degree $5$ and $6$, can be written as
\begin{equation}\label{eq:4.1}
\omega = \tr \left(\ud \sigma \ud c + \ud \rho \ud \phi \right).
\end{equation}
We also have a homological vector field $Q$ of degree $1$ that in the same coordinates reads
\begin{equation}\label{eq:4.2}
\begin{aligned}
Q c &=  \tfrac{1}{2} [c,c] - \phi, \\
Q \phi &= [c,\phi], \\
Q \rho  &=  [c,\rho] - \sigma , \\
Q \sigma &= [c,\sigma] - [\phi,\rho].
\end{aligned}
\end{equation}
One can check that $Q$ preserves $\omega$ and the corresponding hamiltonian of degree $8$ is
\begin{equation}\label{eq:4.3}
\Theta = \tr \left\{  \tfrac{1}{2}\sigma[c,c] - \sigma \phi + \rho[c,\phi] \right\}.
\end{equation}

With these remarks noted, the target manifold is postulated to be the graded manifold $\Ccal(\tau_{[1]}^* P) \times_{T[1]M} \ad(\tau_{[1]}^* P)[2] \times_{T[1]M}  \ad(\tau_{[1]}^* P)[5] \times_{T[1]M}  \ad(\tau_{[1]}^* P)[6]$, where the symbol $\times_{T[1]M}$ stands for the fiber product over $T[1]M$. From the foregoing, it is clear that when this graded manifold is viewed as a bundle over $T[1]M$, its global sections take values in the graded manifold $T[4](\gfrak[1] \oplus \gfrak[2])$. Thus, as expected, our chosen target manifold acquires the structure of a graded symplectic manifold of degree $7$. In fact, if we keep on denoting by $c$, $\phi$, $\rho$ and $\sigma$ the coordinates of each of the factors with degrees $1$, $2$, $5$ and $6$, respectively, then the formulas for the symplectic form of degree $7$, the homological vector field of degree $1$, and the hamiltonian of degree $8$, are all given by the same expressions as \eqref{eq:4.1}, \eqref{eq:4.2} and \eqref{eq:4.3}. 

Moving forward, the next crucial step is to pinpoint the superfield selected for consideration in the AKSZ procedure. In the current scenario, this is taken as a global section $\Phibm \colon T[1]M \to \Ccal(\tau_{[1]}^* P) \times_{T[1]M} \ad(\tau_{[1]}^* P)[2] \times_{T[1]M}  \ad(\tau_{[1]}^* P)[5] \times_{T[1]M}  \ad(\tau_{[1]}^* P)[6]$ of our target manifold. Of course, such a global section may be viewed as a quadruple $\Phibm = (\cbm,\phibm,\rhobm,\sigmabm)$, where $\cbm$ is a generalised connection on $\tau_{[1]}^* P$, $\phibm$ is a global section of $\ad(\tau_{[1]}^* P)[2]$, $\rhobm$ is a global section of $\ad(\tau_{[1]}^* P)[5]$, and $\sigmabm$ is a global section of $\ad(\tau_{[1]}^* P)[6]$. Each of these superfield entries can be further decomposed as
\begin{equation}\label{eq:4.4}
\begin{alignedat}{3}
\cbm &= \sum_{k =0}^{8} c^{(k)}, &\quad \phibm &= \sum_{k =0}^{8} \phi^{(k)}, \\
\rhobm &= \sum_{k =0}^{8} \rho^{(k)}, &\quad \sigmabm &= \sum_{k =0}^{8} \sigma^{(k)},
\end{alignedat}
\end{equation}
in which the $c^{(k)}$, $\phi^{(k)}$, $\rho^{(k)}$ and $\sigma^{(k)}$ are elements of $\Omega^{k}(M,\ad(P))$, except for $c^{(1)}$ which is an ordinary connection on $P$. The ghost number of these various component fields is given by the degree of the superfield entry they appear in minus its degree as a differential form. This means, in particular, that $c^{(1)}$, $\phi^{(2)}$, $\rho^{(5)}$ and $\sigma^{(6)}$ comprise the classical fields, while $c^{(0)}$, $\phi^{(0)}$, $\phi^{(1)}$, $\rho^{(k)}$ for $0 \leq k \leq 4$, and $\sigma^{(k)}$ for $0 \leq k \leq 5$, comprise the ghost fields. The remaining constituents are the antifields, leading to the relationships $c^{(k)} = \sigma^{(8-k)+}$ for $2 \leq k \leq 8$, $\phi^{(k)} = \rho^{(8-k)+}$ for $4 \leq k \leq 8$, $\sigma^{(k)} = c^{(8-k)+}$ for $k = 7,8$, and $\rho^{(k)} = \phi^{(8-k)+}$ for $k = 6,7,8$. For convenience and looking ahead to what follows, we shall henceforth designate $c^{(0)}$ as $c$, $c^{(1)}$ as $\omega$,\footnote{We hope that the use of the same notation for the component field $c^{(1)}$ and for the canonical symplectic form on $T[4](\gfrak[1] \oplus \gfrak[2])$ will not lead to any confusion.} $\phi^{(0)}$ as $\phi$, $\phi^{(1)}$ as $\psi$, and $\phi^{(2)}$ as $G$.

It is now quite straightforward to derive the Batalin-Vilkovisky action together with the BRST transformations. To begin, let us specify the odd symplectic form $\Omega$ on the space of superfields $\Phibm = (\cbm, \phibm,\rhobm,\sigmabm)$. Drawing upon \eqref{eq:2.11} and taking into consideration \eqref{eq:4.1}, we have
\begin{equation}\label{eq:4.5}
\Omega = \int_{T[1]M} \tr \left( \uud \sigmabm \uud \cbm + \uud \rhobm \uud \phibm \right).
\end{equation}
Next, recall from \S\ref{sec:2.3} that the AKSZ algorithm establishes that Batalin-Vilkovisky action $S$ is determined as the hamiltonian function of the BRST operator $\delta$, the latter being characterised as a homological vector field with ghost number $1$, induced by $D$ and $Q$, and acting on the space of superfields. Referring back to \eqref{eq:2.12}, and noting \eqref{eq:4.3}, this reads
\begin{equation}\label{eq:4.6}
S = \int_{T[1]M} \tr \left\{ \sigmabm D \cbm + \rhobm D \phibm + \tfrac{1}{2} \sigmabm [\cbm,\cbm] -  \sigmabm \phibm + \rhobm [\cbm,\phibm]  \right\}.
\end{equation}
Also, the formula \eqref{eq:2.14}, taken in conjunction with \eqref{eq:4.2}, yields the following expression for the BRST transformation rules:
\begin{equation}\label{eq:4.7}
\begin{aligned}
\delta \cbm &= D \cbm +   \tfrac{1}{2} [\cbm,\cbm] - \phibm, \\
\delta \phibm &= D \phibm + [\cbm,\phibm], \\
\delta \rhobm  &= D \rhobm + [\cbm,\rhobm] - \sigmabm , \\
\delta \sigmabm &= D\sigmabm + [\cbm,\sigmabm] - [\phibm,\rhobm].
\end{aligned}
\end{equation}
As remarked in \S\ref{sec:2.3}, the nilpotency of the BRST operator $\delta$ may be reinterpreted as a statement that $S$ satisfies the classical master equation.

We now wish to obtain expressions for the Batalin-Vilkovisky action and the BRST operator in terms of the classical fields, the ghost fields and the antifields appearing in the decomposition \eqref{eq:4.4}. In this regard, as each of these fields carries a natural bigrading, with one grading representing the usual differential form degree and the other corresponding to the ghost number, it becomes necessary for us to adopt a sign convention for their commutation. This is spelt out in appendix~\ref{app:B}. Using these conventions, and performing the $\ud^n \theta$ integration, the full Batalin-Vilkovisky action is
\begin{equation}\label{eq:4.8}
\begin{aligned}
S = \int_M \tr \bigg\{ &c^+ \wedge \left( \tfrac{1}{2}[c,c] - \phi \right) + \omega^{+} \wedge (\ud^{\omega}c - \psi) + \sigma^{(6)} \wedge (F_{\omega} - G + [c,  \sigma^{(6)+} ]) \\
& + \sum_{k=0}^{5} \sigma^{(k)} \wedge \bigg(  \ud^{\omega} \sigma^{(k+1)+} - \rho^{(k)+} + [c,\sigma^{(k)+}] + \tfrac{1}{2} \sum_{i=2}^{6-k} [\sigma^{(8-i)+},\sigma^{(i+k)+}] \bigg) \\
& + \phi^{+} \wedge [c,\phi] + \psi^{+} \wedge (\ud^{\omega} \phi + [c,\psi]) + G^{+} \wedge (\ud^{\omega} \psi + [c,G] - [\phi,\sigma^{(6)+}]) \\ 
& +  \rho^{(5)} \wedge (\ud^{\omega}G + [c,\rho^{(5)+}] - [\phi,\sigma^{(5)+}] - [\psi,\sigma^{(6)+}] ) \\
& + \sum_{k=0}^{4} \rho^{(k)} \wedge \bigg( \ud^{\omega}  \rho^{(k+1)+} + [c, \rho^{(k)+}] - \sum_{i=2}^{8-k} [\rho^{(i+k)+},\sigma^{(8-i)+}] \bigg) \bigg\}.
\end{aligned}
\end{equation}
Accordingly it is found that the BRST transformation rules for the fields are
\begin{equation}\label{eq:4.9}
\begin{gathered}
\delta c =  \tfrac{1}{2}[c,c] - \phi, \quad \delta \omega = \ud^{\omega}c - \psi, \quad \delta \phi = [c,\phi], \\
\delta \psi = \ud^{\omega}\phi + [c,\psi],  \quad \delta G = \ud^{\omega}\psi + [c,G] - [\phi,\sigma^{(6)+}], \\
\delta \rho^{(k)} =  \ud^{\omega} \rho^{(k-1)} - \sigma^{(k)} + [c,\rho^{(k)}] + \sum_{i=k}^{6} [\rho^{(i)+}, \rho^{(i-k)}], \\
\delta \sigma^{(k)} = \ud^{\omega} \sigma^{(k-1)} + [c, \sigma^{(k)}] - [\phi, \rho^{(k)}] - [\psi, \rho^{(k-1)}] + \sum_{i=k}^{6}\left\{ [\sigma^{(i)+}, \sigma^{(i-k)}] - [\rho^{(i)+},\rho^{(i-k)}]\right\},
\end{gathered}
\end{equation}
and, for the antifields, 
\begin{equation}
\begin{gathered}
\delta c^{+} = \ud^{\omega} \omega^+ + [c,c^{+}] - [\phi,\phi^+] - [\psi,\psi^+] + \sum_{i=0}^{6} \left\{ [\sigma^{(i)+}, \sigma^{i}] - [\rho^{(i)+},\rho^{(i)}]\right\} , \\
\delta \omega^{+} = \ud^{\omega} \sigma^{(6)} + [c,\omega^{+}] - [\phi,\psi^+] - [\psi,G^+] + \sum_{i=1}^{6} \left\{ [\sigma^{(i)+}, \sigma^{i-1}] - [\rho^{(i)+},\rho^{(i-1)}]\right\},\\
\delta \phi^{+} =  \ud^{\omega} \psi^{+} - c^+ + [c,\phi^+] + \sum_{i=0}^{6}  [\rho^{(i)+},\rho^{(i)}], \\
\delta \psi^{+} =  \ud^{\omega} G^{+} - \omega^+ + [c,\psi^+] + \sum_{i=1}^{6}  [\rho^{(i)+},\rho^{(i-1)}] , \\
\delta G^{+} =  \ud^{\omega} \rho^{(5)} - \sigma^{(6)} + [c,G^+] + \sum_{i=2}^{6}  [\rho^{(i)+},\rho^{(i-2)}]  , \\
\delta \rho^{(k)+} = \ud^{\omega} \rho^{(k+1)+} + [c,\rho^{(k)+}]  - \sum_{i=2}^{8-k} [\rho^{(i+k)+},\sigma^{(8-i)+}] , \\
\delta \sigma^{(k)+} =  \ud^{\omega} \sigma^{(k+1)+} - \rho^{(k)+}  + [c,\sigma^{(k)+}] + \tfrac{1}{2} \sum_{i=2}^{6-k}[\sigma^{(8-i)+}, \sigma^{(i+k)+}].
\end{gathered}
\end{equation}
It should be pointed out that the structure of the transformations of the sector comprising the fields $c$, $\omega$, $\phi$ and $\psi$ is formally identical to that of the BRST transformations of the fields in the ``geometrical'' sector of Donaldson-Witten theory (see \cite{bonechi2020equivariant,bonechi2023towards,moshayedi20224}, though the terminology is taken from \cite{birmingham1991topological}).

\subsection{Gauge-fixed action}
Our purpose now is to discuss how the action of the topological quantum field theory for $\Spin(7)$-instantons under consideration is recovered from the Batalin-Vilkovisky action after gauge fixing. For this, as we have explained in \S\ref{sec:2.3}, we need to make a judicious choice of a gauge-fixing fermion. 

To start, let us fix an arbitrary metric on $M$, and subsequent to this, let us take the gauge-fixing conditions to be $\pi_7(F_{\omega})$ and $\ud^{\omega}{\ast}\psi$. We introduce therefore a set of antighosts  $\chi \in \Omega^{2}_{7}(M,\ad(P))$ and $\bar{\phi} \in \Omega^{0}(M,\ad(P))$ which have respectively ghost numbers $-1$ and $-2$, and a set of multiplier fields $H \in \Omega^{2}_7(M,\ad(P))$ and $\lambda \in \Omega^{0}(M,\ad(P))$ which have respectively ghost numbers $0$ and $-1$. Naturally the required BRST transformations of these antighosts and multiplier fields are
\begin{equation}\label{eq:4.11}
\begin{alignedat}{3}
\delta \chi  &= H, &\quad \delta H &= 0, \\
\delta \bar{\phi} &= \lambda, &\quad  \delta \lambda &= 0.
\end{alignedat}
\end{equation}
We must also include antifields $\chi^+ \in \Omega^{6}(M,\ad(P))$ and $\bar{\phi}^{+} \in \Omega^{8}(M,\ad(P))$, one for each antighost. The term to be added to the Batalin-Vilkovisky action \eqref{eq:4.8} is then
\begin{equation}\label{eq:4.12}
\int_M \tr \left\{\chi^{+} \wedge H  + \bar{\phi}^{+} \wedge \lambda  \right\}.
\end{equation}
However, in order to match the BRST transformations for the antighosts and multiplier fields featuring in \cite{A-O-S1997} and \cite{B-K-S1998}, we ought to bring forth the following field redefinitions:
\begin{equation}
\begin{aligned}
B &=  [c,\chi] - H, \\
\eta &=  [c,\bar{\phi}] - \lambda.
\end{aligned}
\end{equation} 
Obviously, these redefinitions do not affect any of the above considerations. With these adjustments, the BRST transformations \eqref{eq:4.11} can be brought into the form 
\begin{equation}\label{eq:4.14}
\begin{alignedat}{3}
\delta \chi  &=  [c,\chi] - B, &\quad   \delta B &=  [c,B] - [\phi,\chi], \\
\delta \bar{\phi} &= [c,\bar{\phi}]- \eta, &\quad  \delta \eta &=  [c,\eta] - [\phi,\bar{\phi}],
\end{alignedat}
\end{equation}
while the augmented action can be expressed as
\begin{equation}\label{eq:4.15}
S + \int_M \tr \left\{\chi^{+} \wedge ([c,\chi] - B)  + \bar{\phi}^{+} \wedge ([c,\bar{\phi}] - \eta) \right\}.
\end{equation}
To complete the story, we must choose the gauge-fixing fermion. This is taken to be
\begin{equation}
\Psi = \int_M \tr \left\{   \tfrac{1}{2}  \chi \wedge {\ast} B - \chi \wedge {\ast} \pi_7(F_{\omega}) + \bar{\phi} \wedge \ud^{\omega}{\ast}\psi + \tfrac{1}{2} c [\chi,{\ast}\chi] \right\}.
\end{equation}
The gauge-fixed action is constructed from \eqref{eq:4.15} by restricting the antifields to lie on the lagrangian submanifold specified by $\Psi$. This restriction results in the conditions
\begin{equation}\label{eq:4.17}
\begin{gathered}
\begin{alignedat}{1}
& c^{+}  =  \frac{\delta \Psi}{\delta c} =  \tfrac{1}{2} [\chi,{\ast}\chi], \quad   \omega^+ =  \frac{\delta \Psi}{\delta \omega} = \ud^{\omega} {\ast}\chi  + [\bar{\phi},{\ast}\psi], \quad \psi^{+}   = \frac{\delta \Psi}{\delta \psi} = {\ast}\ud^{\omega} \bar{\phi}, 
\end{alignedat} \\
\begin{alignedat}{1}
& \phi^{+} = \frac{\delta \Psi}{\delta \phi} = 0, \quad  G^{+} = \frac{\delta \Psi}{\delta G} = 0, \quad \rho^{(k)+} = \frac{\delta \Psi}{\delta \rho^{(k)}} = 0, \quad \sigma^{(k)+} = \frac{\delta \Psi}{\delta \sigma^{(k)}} = 0,
\end{alignedat} \\
\begin{alignedat}{1}
&  \chi^{+} = \frac{\delta \Psi}{\delta \chi} = {\ast}\pi_7(F_{\omega}) - \tfrac{1}{2} {\ast} B -  {\ast}[c,\chi], \quad   \bar{\phi}^{+} = \frac{\delta \Psi}{\delta \bar{\phi}} = \ud^{\omega} {\ast} \psi.
\end{alignedat} 
\end{gathered}
\end{equation}
In deriving these equalities we make use of the formulas and sign conventions detailed in appendix~\ref{app:B}. After substituting \eqref{eq:4.17} into \eqref{eq:4.15} and some rearrangement, the gauge-fixed action becomes
\begin{equation}\label{eq:4.18}
\begin{aligned}
I_{\Psi} = \int_M \tr \big\{  &\tfrac{1}{4} [c,c] [\chi,{\ast}\chi]   - \tfrac{1}{2} \phi [\chi,{\ast}\chi]  + \ud^{\omega} {\ast}\chi \wedge \ud^{\omega} c - \ud^{\omega} {\ast}\chi \wedge \psi + [\bar{\phi},{\ast}\psi] \wedge \ud^{\omega} c \\
& -[\bar{\phi},{\ast}\psi] \wedge \psi + {\ast} \ud^{\omega} \bar{\phi} \wedge \ud^{\omega} \phi + {\ast} \ud^{\omega} \bar{\phi} \wedge [c,\psi] - \rho^{(5)} \wedge \ud^{\omega}G + \sigma^{(6)} \wedge (F_{\omega} - G) \\
& + {\ast} \pi_7(F_{\omega}) \wedge B  -\tfrac{1}{2} {\ast}B \wedge B + \tfrac{3}{2} {\ast} [c,\chi] \wedge B - {\ast} \pi_7(F_{\omega}) \wedge [c,\chi]  - {\ast}[c,\chi] \wedge [c,\chi]  \\
& + \ud^{\omega} {\ast}\psi \wedge [c,\bar{\phi}] - \eta \ud^{\omega} {\ast}\psi   \big\}.
\end{aligned}
\end{equation}
From this we can read off the fact that the fields $\rho^{(5)}$, $\sigma^{(6)}$ and $B$ only enter algebraically, which indicates they may be regarded as auxiliary fields. These fields can therefore be eliminated through the use of their own equations of motion. For the fields $\rho^{(5)}$ and $\sigma^{(6)}$, the equations of motion are $\ud^\omega G=0$ and $F_{\omega} - G = 0$, respectively, both of which are jointly satisfied by virtue of the Bianchi identity. As for the field $B$, its equation of motion gives 
\begin{equation}\label{eq:4.19}
B = \pi_7(F_{\omega}) + \tfrac{3}{2}[c,\chi].
\end{equation}
It should also be noted that in the gauge-fixed action \eqref{eq:4.18}, several simplifications occur. Indeed, employing calculations similar to those presented in \S\ref{sec:3.2} and adhering again to the sign conventions outlined in appendix~\ref{app:B}, one can demonstrate the validity of the following identities:
\begin{equation}
\begin{gathered}
\int_M \tr \left(\ud^{\omega}{\ast} \chi \wedge \psi \right) = - \int_M \tr \left(\ud^{\omega} \psi \wedge {\ast} \chi \right), \\
\int_M \tr \left( [\bar{\phi},{\ast}\psi] \wedge \psi \right) = - \int_M \tr \left( \bar{\phi}[\psi,{\ast}\psi]\right),\\
\int_M \tr \left( {\ast}\ud^{\omega}\bar{\phi} \wedge \ud^{\omega} \phi \right) =  \int_M \tr \left( \phi {\ast} \Delta_0^{\omega} \bar{\phi} \right),\\
\int_M \tr \left({\ast}[c,\chi] \wedge [c,\chi] \right) = - \int_M \tr \left( \tfrac{1}{2} [c,c]  [\chi,{\ast}\chi] \right), \\
\int_M \tr \left(  [\bar{\phi},{\ast}\psi] \wedge \ud^{\omega} c + {\ast} \ud^{\omega} \bar{\phi} \wedge [c,\psi] + \ud^{\omega} {\ast}\psi \wedge [c,\bar{\phi}] \right) = 0.
\end{gathered}
\end{equation}
Bearing in mind all the foregoing, and subsequent to regrouping terms, the gauge-fixed action emerges as
\begin{equation}\label{eq:4.21}
\begin{aligned}
I_{\Psi} = \int_M \tr &\left\{  \tfrac{1}{2} \pi_7(F_{\omega}) \wedge {\ast} \pi_7(F_{\omega}) + \ud^{\omega} \psi \wedge {\ast} \chi - \tfrac{1}{2}\phi[\chi,{\ast}\chi] + \bar{\phi}[\psi,{\ast}\psi] + \phi {\ast} \Delta_0^{\omega} \bar{\phi} - \eta \ud^{\omega} {\ast}\psi  \right. \\
&\left. \!\!\phantom{\tfrac{1}{2}} + \tfrac{1}{2} [c,\chi] \wedge {\ast} \pi_7(F_{\omega}) + \ud^{\omega} {\ast}\chi \wedge \ud^{\omega} c + \tfrac{3}{16} [c,c]  [\chi,{\ast}\chi]  \right\}.
\end{aligned}
\end{equation}
After rewriting the first term using the identity \eqref{eq:3.10}, a straightforward look reveals that the first line of this expression matches all the terms in the action \eqref{eq:3.27}, differing only in the coefficients. We are thus led to conclude that the gauge-fixed action \eqref{eq:4.21} is equivalent to the one obtained via the Mathai–Quillen formalism, up to additional ghost interactions.

The residual BRST transformations correspond exactly to the BRST transformation rules induced on the fields \eqref{eq:4.9} upon imposing the gauge fixing conditions \eqref{eq:4.17}, together with those of \eqref{eq:4.14} after imposing the equation of motion \eqref{eq:4.19}. They read
\begin{equation} \label{eq:4.22}
\begin{gathered}
\delta c = \tfrac{1}{2}[c,c] - \phi, \quad  \delta \omega = \ud^{\omega} c - \psi, \quad 
\delta \phi = [c,\phi],  \quad \delta \psi = \ud^{\omega} \phi + [c,\psi], \\
\delta \chi = - \pi_{7}(F_{\omega}) - \tfrac{1}{2}[c,\chi], \quad \delta \bar{\phi} = [c,\bar{\phi}] - \eta, \quad \delta \eta = [c,\eta] - [\phi,\bar{\phi}]. 
\end{gathered}
\end{equation}
We remark, however, that these transformations are nilpotent only ``on-shell''. Indeed, $\delta^2$ vanishes identically for all fields except $\chi$, for which the equations of motion must be used. This turns out to be a consequence of having eliminated the auxiliary field $B$, which has partially broken the nilpotency of the BRST operator.

\subsection{Observables}
Having derived the gauge-fixed action, we now focus on the observables of the theory. As explained in \S\ref{sec:2.3}, this involves choosing appropriate representatives in the cohomology of the Hamiltonian vector field $Q$.

To proceed, we need to understand what functions on the graded manifold $T[4](\mathfrak{g}[1] \oplus \mathfrak{g}[2])$ look like. In order to describe them explicitly, it is convenient to choose a basis ${e^a}$ for the Lie algebra $\mathfrak{g}$, which induces components $\rho^a$ and $\sigma^a$ for the fiber coordinates. A generic function $f$ on this graded manifold can then be written as
\begin{equation}
f = \tfrac{1}{k!l!} f_{a_1 \cdots a_k b_1 \cdots b_l}(c,\phi)  \rho^{a_1} \cdots \rho^{a_k} \sigma^{b_1} \cdots \sigma^{b_l},
\end{equation}
where $f_{a_1 \cdots a_k b_1 \cdots b_l}$ is an arbitrary function of the base coordinates $c$ and $\phi$, antisymmetric in the indices $a_i$ and symmetric in the indices $b_i$. This directly implies that the pullback of any such $f$ along any superfield $\Phibm = (\cbm, \phibm, \rhobm, \sigmabm)$ takes the form
\begin{equation}
\Phibm^* f = \tfrac{1}{k!l!} f_{a_1 \cdots a_k b_1 \cdots b_l}(\cbm,\phibm)  \rhobm^{a_1} \cdots \rhobm^{a_k} \sigmabm^{b_1} \cdots \sigmabm^{b_l}.
\end{equation}
We may then decompose $\Phibm^* f$ according to the form degree on $M$, making use of the decompositions introduced in \eqref{eq:4.4}. When $f$ is taken to be $Q$-closed, this decomposition gives rise to a collection of eight relations of the form \eqref{eq:2.22}, organized by this form degree on $M$. A complete set of classical observables can thus be obtained by integrating over the appropriate homology cycles on $M$ for each term in the decomposition.

Our particular interest lies in reproducing the hierarchy of observables that we found in \S\ref{sec:3.3}. To achieve this, we shall explicitly restrict our field configurations to those constrained on the Lagrangian submanifold defined by the gauge-fixing fermion $\Psi$ introduced in the preceding subsection. Having thus restricted our field configurations, and with the auxiliary fields integrated out, the relevant superfields reduce to the combinations
\begin{align}
\cbm &= c + \omega, \\
\phibm &= \phi + \psi + F_{\omega}. \label{eq:4.26}
\end{align}
In this form, we may loosely think of $\phibm$ as playing the role of a ``curvature multiplet'' associated to the ``connection multiplet'' $\cbm$. Now, borrowed from the discussion in \S\ref{sec:3.3}, we consider the function $f_n$ on the graded manifold $T[4](\gfrak[1] \oplus \gfrak[2])$ defined by
\begin{equation}
f_n = \tr \phi^n.
\end{equation}
Owing to the explicit form of the Hamiltonian vector field $Q$ displayed in \eqref{eq:4.2}, one readily checks that $f_n$ is indeed $Q$-closed. From this, we immediately identify
\begin{equation} \label{eq:4.28}
\Wbm_{\!\!\! n}  = \Phibm^* f_n = \tr \phibm^n
\end{equation}
as a quantity that gives rise to classical observables. Let us make the decomposition of this quantity with respect to the form degree on $M$, by substituting the component field expansion \eqref{eq:4.26} of $\phibm$ into \eqref{eq:4.28}. One finds
\begin{equation} \label{eq:4.29}
\Wbm_{\!\!\! n} = \sum_{k=0}^{8} W_n^{(k)},
\end{equation}
where the $W_n^{(k)}$ are the $k$-forms on $M$, the first five of which are given explicitly by
\begin{equation}
\begin{gathered}
W_n^{(0)} = \tr \phi^n, \quad  W_n^{(1)} = \tbinom{n}{1} \tr \left( \phi^{n-1}\wedge \psi \right), \quad  W_n^{(2)} = \tbinom{n}{1}\tr \left(\phi^{n-1} \wedge F_{\omega} \right) + \tbinom{n}{2} \tr \left( \phi^{n-2} \wedge \psi \wedge \psi \right),  \\
W_n^{(3)} = 2 \tbinom{n}{2} \tr \left( \phi^{n-2} \wedge \psi \wedge F_{\omega}\right) + \tbinom{n}{3} \tr \left( \phi^{n-3} \wedge \psi \wedge \psi \wedge \psi \right),\\
W_n^{(4)} = \tbinom{n}{2} \tr \left( \phi^{n-2} \wedge F_{\omega} \wedge F_{\omega} \right) + 3\tbinom{n}{3} \tr \left( \phi^{n-3} \wedge \psi \wedge \psi \wedge F_{\omega}\right) + \tbinom{n}{4} \tr \left( \phi^{n-4} \wedge \psi \wedge \psi \wedge \psi \wedge \psi  \right).
\end{gathered}
\end{equation}
In the case $n = 2$, these quantities reduce, up to an overall factor of $\frac{1}{2}$, to those constructed in \S\ref{sec:3.3}, provided they are reinterpreted appropriately.\footnote{Each term in these quantities can be interpreted as a homogeneous component of a $\Gcal^*$-equivariant differential form on the product $M \times \Acal^*$.} As for the remaining four $k$-forms, their expressions are
\begin{equation}
\begin{aligned}
W_n^{(5)} &= 3\tbinom{n}{3} \tr \left(\phi^{n-3} \wedge \psi \wedge F_{\omega}^2 \right) + \tbinom{n}{4} \tr \left(\phi^{n-4} \wedge \psi^3 \wedge F_{\omega} \right) +   \tbinom{n}{6} \tr \left(\phi^{n-5} \wedge \psi^5 \right),\\
W_n^{(6)} &= \tbinom{n}{3} \tr \left(\phi^{n-3} \wedge F_{\omega}^3 \right) + 6\tbinom{n}{4} \tr \left(\phi^{n-4} \wedge \psi^2 \wedge F_{\omega}^2 \right) + 5\tbinom{n}{5} \tr \left(\phi^{n-5} \wedge \psi^4 \wedge F_{\omega} \right) \\
&\phantom{=}\  + \tbinom{n}{6} \tr \left(\phi^{n-6} \wedge \psi^6 \right),\\
W_n^{(7)} &= 4\tbinom{n}{4} \tr \left(\phi^{n-4}  \wedge \psi \wedge F_{\omega}^3 \right) + 10 \tbinom{n}{5} \tr \left(\phi^{n-5} \wedge   \psi^3 \wedge F_{\omega}^2 \right) + 6 \tbinom{n}{6} \tr \left(\phi^{n-6} \wedge  \psi^5 \wedge F_{\omega} \right) \\
&\phantom{=}\  + \tbinom{n}{7} \tr \left(\phi^{n-7} \wedge \psi^7  \right),\\
W_n^{(8)} &= \tbinom{n}{4} \tr \left( \phi^{n-4} \wedge F_{\omega}^4 \right) + 10 \tbinom{n}{5} \tr \left( \phi^{n-5} \wedge \psi^2 \wedge F_{\omega}^3\right) + 15 \tbinom{n}{6} \tr \left(\phi^{n-6} \wedge \psi ^4  \wedge F_{\omega}^2 \right)\\
&\phantom{=}\  + 7\tbinom{n}{7} \tr \left( \phi^{n-7} \wedge \psi^6 \wedge F_{\omega} \right) + \tbinom{n}{8} \tr \left(\phi^{n-8} \wedge \psi^8 \right),
\end{aligned}
\end{equation}
where we use the shorthands $\psi^{i}$ and $F_{\omega}^{i}$ to denote the wedge products of $\psi$ and $F_{\omega}$ with themselves $i$ times. Building on this, the implication of \eqref{eq:2.22} is that the forms $W_n^{(k)}$ are subject to the following chain of equalities:
\begin{equation}
\delta W_n^{(k)} = \ud W_n^{(k-1)}.
\end{equation}
Here, it is understood that the BRST operator $\delta$ corresponds to the residual one, whose transformation rules are given in \eqref{eq:4.22}. The classical observables are then made up of $\Ocal_n^{(k)}(\gamma) = \int_{\gamma} W_n^{(k)}$ where $\gamma$ is a $k$-homology cycle on $M$. 

Let us now take a step back and examine a general structural aspect of the construction above. As shown in appendix~\ref{app:C}, the cohomology of the homological vector field $Q$ is trivial when acting on functions on the graded manifold $T[4](\gfrak[1] \oplus \gfrak[2])$. Therefore, we can find a function $s_n$ on $T[4](\gfrak[1] \oplus \gfrak[2])$ of total degree $2n-1$ such that
\begin{equation}
f_n = Q s_n.
\end{equation}
Let us define $\Ibm_{\!\! n}$ by
\begin{equation}
\Ibm_{\!\! n} = \Phibm^* s_n,
\end{equation}
and notice that 
\begin{equation}
\delta \Ibm_{\!\! n} = D \Phibm^* s_n +  \Phibm^* Qs_n = D \Phibm^* s_n +  \Phibm^* f_n = D \Ibm_{\!\! n} + \Wbm_{\!\!\! n},
\end{equation}
or, equivalently,
\begin{equation}\label{eq:4.36}
\Wbm_{\!\!\! n} = \delta \Ibm_{\!\! n} - D \Ibm_{\!\! n}.
\end{equation}
Just as we decomposed $\Wbm_{\!\!\! n}$ according to the form degree on $M$, we may similarly expand $\Ibm_{\!\! n}$ as
\begin{equation}
\Ibm_{\!\! n} = \sum_{k = 0}^{7} I_n^{(k)},
\end{equation}
each $I_n^{(k)}$ being a $k$-form on $M$. This, together with \eqref{eq:4.29}, indicates that a component-by-component comparison of \eqref{eq:4.36} results in
\begin{equation}
W_n^{(k)} = \delta I_n^{(k)} - \ud I_n^{(k-1)}.
\end{equation}
Thus, we can assert that $W_n^{(k)}$ is $\delta$-exact, up to a total derivative.

To close, it is worth emphasising that the present framework extends to incorporate the further observables described in §\ref{sec:3.3}, which arise in the case where the $\Spin(7)$-structure $\Omega$ is integrable. Rather than dealing with a function on the graded manifold $T[4](\gfrak[1] \oplus \gfrak[2])$, one instead considers a function on the graded manifold $\Ccal(\tau_{[1]}^* P) \times_{T[1]M} \ad(\tau_{[1]}^* P)[2] \times_{T[1]M} \ad(\tau_{[1]}^* P)[5] \times_{T[1]M} \ad(\tau_{[1]}^* P)[6]$, where multiplication by $\Omega$ is meaningful. The relevant function is then
\begin{equation}
\tilde{f}_2 = \left( \tr  \phi^2 \right) \Omega,
\end{equation}
which turns out to be $Q$-closed, as $\Omega$ is $Q$-closed. As a result, the quantity
\begin{equation} \label{eq:4.40}
\tilde{\Wbm}_{\!\!\! 2}  = \Phibm^* \tilde{f}_2 =  \left( \tr  \phibm^2 \right) \Omega
\end{equation}
can be used to build classical observables. Upon decomposing $\tilde{\Wbm}_{\!\!\! 2} $ with respect to the form degree on $M$, by inserting the expansion \eqref{eq:4.26} for $\phibm$ into \eqref{eq:4.40}, one finds
\begin{equation}
\tilde{\Wbm}_{\!\!\! 2}  = \sum_{k=4}^{8} \tilde{W}_2^{(k)},
\end{equation}
where each of the $k$-forms $\tilde{W}_2^{(k)}$ is obtained as
\begin{equation}
\tilde{W}_2^{(k)} = W_2^{(k-4)} \wedge \Omega.
\end{equation}
These quantities precisely reproduce those described in the final part of \S\ref{sec:3.3}, which, as noted there, were originally proposed in \cite{A-O-S1997} and \cite{B-K-S1998}.

\subsection{Chern-Simons formulation}
In this final subsection, we show that the Batalin–Vilkovisky action admits a reformulation in terms of a formal Chern–Simons theory, in a way closely analogous to that offered by Costello in Section 2 of Chapter 6 of \cite{costello2011renormalization} for Yang–Mills theory. The perspective we adopt is motivated by certain ideas from \cite{bonechi2023towards}, developed in the context of Donaldson–Witten theory, though adapted here to fit our framework and goals.

The procedure we follow naturally splits into two steps. The first step consists in casting the Batalin–Vilkovisky action in a form reminiscent of the corresponding action underlying Donaldson–Witten theory. To start off, we take the product $T[1]M \times \RR[-4]$ as our primary geometric datum, and set $\nu$ to denote the extra odd coordinate on $\RR[-4]$. We naturally extend the homological vector field $D$ on $T[1]M$ to this product, and keep using the same letter to refer to it. We also write $\tau^{\prime}_{[1]} \colon T[1]M \times \RR[-4] \to M$ for the bundle projection of $T[1]M \times \RR[-4]$ onto $M$. As before, we may then consider the pullback bundle $\tau^{\prime *}_{[1]}P$ of $P$ along $\tau^{\prime}_{[1]}$, which defines a principal bundle over $T[1]M \times \RR[-4]$ with structure group $G$, together with its associated adjoint bundle $\ad(\tau^{\prime *}_{[1]}P) = \tau^{\prime *}_{[1]}P \times_{\ad} \gfrak$. The secondary geometric datum is taken to be the graded manifold $\Ccal(\tau^{\prime *}_{[1]}P) \times_{T[1]M \times \RR[-4]} \ad(\tau^{\prime *}_{[1]}P)[2]$, viewed as a bundle over $T[1]M \times \RR[-4]$. Its sections necessarily take values in the graded manifold $\gfrak[1] \oplus \gfrak[2]$. According to \cite{bonechi2023towards}, the graded manifold $\Ccal(\tau^{\prime *}_{[1]}P) \times_{T[1]M \times \RR[-4]} \ad(\tau^{\prime *}_{[1]}P)[2]$ carries a symplectic form of degree $3$, along with a homological vector field of degree $1$ that preserves it, which is generated by a hamiltonian of degree $4$. If we denote by $c'$ and $\phi'$ the coordinates of each factor, with degrees $1$ and $2$ respectively, the symplectic form is that given by
\begin{equation}\label{eq:4.43}
\omega = \tr \left( \ud c' \ud \phi' \right).
\end{equation}
As for the homological vector field, in these same coordinates, it takes the form
\begin{equation}\label{eq:4.44}
\begin{aligned}
Q c' &= \tfrac{1}{2}[c',c'] - \phi', \\
Q \phi' &= [c',\phi'],
\end{aligned}
\end{equation}
while the hamiltonian is expressed as
\begin{equation}\label{eq:4.45}
\Theta = \tr \left\{ \tfrac{1}{2} \phi' [c',c'] - \phi' \phi'  \right\}. 
\end{equation}
Now, the crucial point of all this is that, if we are given a global section of the bundle $\Ccal(\tau_{[1]}^* P) \times_{T[1]M} \ad(\tau_{[1]}^* P)[2] \times_{T[1]M} \ad(\tau_{[1]}^* P)[5] \times_{T[1]M} \ad(\tau_{[1]}^* P)[6]$ over $T[1]M$, which, as we saw, is specified by the components $\cbm$, $\phibm$, $\rhobm$, and $\sigmabm$, then we can always induce a global section of the bundle $\Ccal(\tau_{[1]}^{\prime *} P) \times_{T[1]M} \ad(\tau_{[1]}^{\prime *} P)[2]$ over $T[1]M \times \RR[-4]$, which can be understood as consisting of a generalised connection $\cbm’$ on $\tau_{[1]}^{\prime *} P$ and a global section $\phi'$ of $\ad(\tau_{[1]}^{\prime *} P)$, determined by the prescription
\begin{equation}\label{eq:4.46}
\begin{aligned}
\cbm’ &= \cbm + \nu \rhobm, \\
\phibm’ &= \phibm + \nu \sigmabm.
\end{aligned}
\end{equation}
Hence, within the AKSZ formalism, our theory may be reinterpreted as one where the source manifold is $T[1]M \times \RR[-4]$ and the target manifold is $\Ccal(\tau_{[1]}^{\prime *} P) \times_{T[1]M} \ad(\tau_{[1]}^{\prime *} P)[2]$, with the superfields described as global sections of the latter, whose components are given by \eqref{eq:4.46}. It is then not difficult to convince oneself that, in this new setup, the odd symplectic form $\Omega$ given in \eqref{eq:4.5} can be expressed as
\begin{equation}\label{eq:4.47}
\Omega = \int_{T[1]M \times \RR[-4]} \tr \left(\uud \cbm’ \uud \phibm’ \right),
\end{equation}
in full agreement with what one would expect from \eqref{eq:4.43}. Likewise, the Batalin–Vilkovisky action $S$ given in \eqref{eq:4.6} can be rewritten as
\begin{equation}\label{eq:4.48}
S = \int_{T[1]M \times \RR[-4]} \tr \left\{ \phibm’ D \cbm’ - \tfrac{1}{2} \phibm’ \phibm’ + \tfrac{1}{2} \phibm’ [\cbm’,\cbm’] \right\},
\end{equation}
and the BRST transformation rules reduce to
\begin{equation}\label{eq:4.49}
\begin{aligned}
\delta \cbm’ &= D \cbm’ - \phibm’ + \tfrac{1}{2} [\cbm’,\cbm’], \\
\delta \phibm’ &= D \phibm’ + [\cbm’,\phibm’],
\end{aligned}
\end{equation}
which is consistent with \eqref{eq:4.44} and \eqref{eq:4.45}. Thus, the geometric parallel between our construction and Donaldson–Witten theory becomes explicit.

The second step consists in adjusting the procedure presented in \cite{bonechi2023towards}, so as to transform the Batalin–Vilkovisky action in \eqref{eq:4.48} into one of Chern–Simons type. We must therefore follow a line of reasoning very similar to that discussed above. The first move is to enlarge what we have called the primary geometric datum to the product $T[1]M \times \RR[-4] \times \RR[-1]$, in which we use $\xi$ to denote the odd coordinate on the new factor $\RR[-1]$. In this product, the homological vector field we will consider is $D - \frac{\partial}{\partial \xi}$. We will also denote by $\tau_{[1]}^{\prime\prime} \colon T[1]M \times \RR[-4] \times \RR[-1] \to T[1]M$ the projection onto $T[1]M$, and consider again the pullback bundle $\tau_{[1]}^{\prime\prime *}P$, along with its associated adjoint bundle $\ad(\tau_{[1]}^{\prime\prime *}P) = \tau_{[1]}^{\prime\prime *}P \times_{\ad} \gfrak$. As for the secondary geometric datum, it is now simply taken to be the graded manifold $\Ccal(\tau_{[1]}^{\prime\prime *}P)$, viewed as a bundle over $T[1]M \times \RR[-4] \times \RR[-1]$, whose sections take values in the graded manifold $\gfrak[1]$. This graded manifold, as required, supports a symplectic form of degree $2$, a homological vector field, and a hamiltonian of degree $3$, which, with its generic coordinate denoted by $A$, can be written, respectively, as
\begin{equation}\label{eq:4.50}
\omega = \tfrac{1}{2} \tr \left( \ud A \ud A\right), \quad \delta A =  \tfrac{1}{2}  [A,A], \quad 
\Theta = \tfrac{1}{6} \tr \left\{ A [A,A]\right\}.
\end{equation}
And the essential point here is that, given a global section of the bundle $ \Ccal(\tau_{[1]}^{\prime *} P) \times_{T[1]M} \ad(\tau_{[1]}^{\prime *} P)[2] $ over $ T[1]M \times \RR[-4] $, determined by a pair $ \cbm' $ and $ \phibm' $ as explained earlier, we can produce a global section of the bundle $ \Ccal(\tau_{[1]}^{\prime\prime *} P) $ over $T[1]M \times \RR[-4] \times \RR[-1]$, which can be regarded as a generalised connection $ \Abm $ on $ \tau_{[1]}^{\prime\prime *} P $, and it is defined through the formula
\begin{equation}\label{eq:4.51}
\Abm = \cbm' + \xi \phibm'.
\end{equation}
Consequently, we can reinterpret our theory once again within the framework of the AKSZ formalism, as one where the source manifold is $T[1]M \times \RR[-4] \times \RR[-1]$, the target manifold is $\Ccal(\tau_{[1]}^{\prime\prime *}P)$, and there is a single superfield characterized as a section of the latter, given by \eqref{eq:4.51}. In terms of this superfield, the odd symplectic form $\Omega$ given in \eqref{eq:4.47} can be represented as
\begin{equation}
\Omega = \tfrac{1}{2} \int_{T[1]M \times \RR[-4] \times \RR[-1]} \tr \left( \uud \Abm \uud \Abm \right),
\end{equation}
whereas the Batalin–Vilkovisky action $S$ given in \eqref{eq:4.48} takes the form
\begin{equation}\label{eq:4.53}
S = \int_{T[1]M \times \RR[-4] \times \RR[-1]} \tr \left\{ \tfrac{1}{2} \Abm \left( D - \frac{\partial}{\partial \xi} \right) \Abm + \tfrac{1}{6} \Abm [\Abm,\Abm]\right\}.
\end{equation}
Furthermore, the BRST transformations in \eqref{eq:4.49} reduce to a single transformation, which is given by
\begin{equation}
\delta \Abm = \tfrac{1}{2} \left( D - \frac{\partial}{\partial \xi} \right) \Abm + \tfrac{1}{2} [\Abm,\Abm]. 
\end{equation}
Notice that these three final expressions are exactly what we would expect to obtain from those in \eqref{eq:4.50} by direct application of the AKSZ construction. We have thus achieved our goal, as the action in \eqref{eq:4.53} has the format of a Chern-Simons action. In fact, using a slightly more abstract language, such as that in \cite{Ben-Shahar:2024dju}, we can say that the theory we have constructed is the formal Chern-Simons theory associated with the differential Frobenius algebra defined by the space of functions of the graded manifold $T[1]M \times \RR[-4] \times \RR[-1]$, equipped with the differential $D - \frac{\partial}{\partial \xi}$ and the non-degenerate pairing $\int_{T[1]M \times \RR[-4] \times \RR[-1]}$.


\appendix
\section{The Cartan model for equivariant cohomology}\label{app:A}
This appendix provides a brief account of the Cartan model for equivariant cohomology in the context of Lie group actions on manifolds. This model can be viewed as a deformation of the de Rham complex that incorporates data from the group action.

Let $M$ be a manifold and let $G$ be a compact, connected Lie group acting smoothly on $M$ on the left. We denote the action by $\sigma \colon G \times M \to M$ and write $\sigma(g,x) = g \cdot x = \sigma_g(x)$. Let $\gfrak$ be the Lie algebra of $G$, $\Pol^{\sbullet}(\gfrak)$ the algebra of  polynomial functions on $\gfrak$ and $\Omega^{\sbullet}(M)$ the algebra of differential forms on $M$. We consider the tensor product $\Pol^{\sbullet}(\gfrak) \otimes \Omega^{\sbullet}(M)$, every element of which is a sum of terms of the form $\alpha = f \otimes \varphi$, where $f \in \Pol^{\sbullet}(\gfrak)$ and $\varphi \in \Omega^{\sbullet}(M)$. These are best thought of as $\Omega^{\sbullet}(M)$-valued polynomial functions on $\gfrak$, so that an element $\alpha = f \otimes \varphi$ acts on $\xi \in \gfrak$ by $\alpha(\xi) = f(\xi)\varphi$. Now note that $\Pol^{\sbullet}(\gfrak) \otimes \Omega^{\sbullet}(M)$ has a natural grading:~it is the differential form grading on $\Omega^{\sbullet}(M)$ plus two times the polynomial grading on $\Pol^{\sbullet}(\gfrak)$. Moreover, the action of $G$ on $M$ together with the adjoint action of $G$ on $\gfrak$ give a natural action of $G$ on $\Pol^{\sbullet}(\gfrak) \otimes \Omega^{\sbullet}(M)$. Explicitly, if $\alpha = f \otimes \varphi$ and $g \in G$, then $g \cdot \alpha$ is the element of $\Pol^{\sbullet}(\gfrak) \otimes \Omega^{\sbullet}(M)$ whose value at any $\xi \in \gfrak$ is
\begin{equation}
(g \cdot \alpha)(\xi) = (g \cdot (f \otimes \varphi))(\xi) = f(\ad_{g^{-1}}\xi) \sigma_{g^{-1}}^*\varphi.
\end{equation}
The algebra of all $G$-invariant elements of $\Pol^{\sbullet}(\gfrak) \otimes \Omega^{\sbullet}(M)$ is denoted $\Omega^{\sbullet}_{G}(M)$ and its elements are called $G$-equivariant differential forms on $M$. We define the Cartan exterior derivative $\ud_G$ on  $\Pol^{\sbullet}(\gfrak) \otimes \Omega^{\sbullet}(M)$ by the formula
\begin{equation}\label{eq:A.2}
(\ud_G \alpha)(\xi) = \ud(\alpha(\xi)) - i_{\xi^{\#}}(\alpha(\xi)),
\end{equation}
for any $\alpha \in \Pol^{\sbullet}(\gfrak) \otimes \Omega^{\sbullet}(M)$ and $\xi \in \gfrak$. Here, $\xi^{\#}$ is the vector field on $M$ defined at each $x \in M$ by 
\begin{equation}\label{eq:A.3}
\xi^{\#}(x) = \frac{\ud}{\ud t} \bigg\vert_{t=0}  \exp(-t \xi) \cdot x,
\end{equation}
and $i_{\xi^{\#}}$ denotes interior product by $\xi^{\#}$. It can be shown that $\ud_G$ raises the degree by one. One also verifies that $\ud_G$ preserves the subalgebra $\Omega^{\sbullet}_G(M)$ of
invariant elements and satisfies, for any $\alpha \in \Pol^{\sbullet}(\gfrak) \otimes \Omega^{\sbullet}(M)$ and any $\xi \in \gfrak$, 
\begin{equation}\label{eq:A.4}
(\ud_G^2(\alpha))(\xi) = - L_{\xi^{\#}}(\alpha(\xi)),
\end{equation}
where $L_{\xi^{\#}}$ is the Lie derivative with respect to $\xi^{\#}$. Since an invariant element $\alpha$ of $\Pol^{\sbullet}(\gfrak) \otimes \Omega^{\sbullet}(M)$ satisfies $L_{\xi^{\#}}(\alpha(\xi))$ for every $\xi \in \gfrak$, we obtain from \eqref{eq:A.4} that $\ud_G^2 = 0$ on $\Omega^{\sbullet}_G(M)$. Thus $\Omega^{\sbullet}_G(M)$ with $\ud_G$ acting on it defines a cochain complex. The cohomology of this complex is called the Cartan model of the $G$-equivariant cohomology of $M$ and is denoted $H^{\sbullet}_G(M)$.

Notice that if $M$ is a single point $\pt$, then every element of $\Omega^{\sbullet}_G(\pt)$ is of the form $f \otimes 1$ for some $f \in \Pol(\gfrak)$ which is invariant under the action of $G$. Each of these is $G$-equivariantly closed, but none is $G$-equivariantly exact, so $H^{\sbullet}_G(\pt)$ is isomorphic to the algebra of $G$-invariant elements of $\Pol^{\sbullet}(\gfrak)$. Notice also that if $G$ is trivial, then so is its Lie algebra $\gfrak$, so there are only constant polynomials on $\gfrak$. Everything is $G$-invariant, so one can identify $\Omega^{\sbullet}_G(M)$ with $\Omega^{\sbullet}(M)$. Furthermore, $\ud_G$ agrees with the ordinary de Rham differential $\ud$, and we conclude that $H^{\sbullet}_G(M)$ is isomorphic to the de Rham cohomology $H^{\sbullet}_{\mathrm{DR}}(M)$.


\section{Conventions}\label{app:B}
In this appendix, we outline our conventions regarding the graded objects used in section~\ref{sec:4} and present the properties necessary for our calculations. The setting we use throughout is as described in that section. One class of graded objects we consider consists of differential forms on $M$ with values in the adjoint bundle $\ad(P)$. These are endowed with a natural bigrading: one grading corresponds to the usual form degree, while the other labels the ghost number. Given one such form $\alpha$, we shall write $\deg(\alpha)$ for the former and $\gh(\alpha)$ for the latter; we also denote their sum, the total degree, by $\lvert \alpha \rvert$. The following are the general properties of these bigraded forms. To begin with, the graded commutator, which was used in section \ref{sec:4}, is defined as
\begin{equation}
[\alpha, \beta] = \alpha \wedge \beta  - (-1)^{\lvert \alpha \rvert \lvert \beta \rvert } \beta \wedge \alpha.
\end{equation}
Thus if $\lvert \alpha \rvert$ or $\lvert \beta \rvert$ is even we obtain the commutator, while if $\lvert \alpha \rvert$ and $\lvert \beta \rvert$ are odd we get the anticommutator. As expected, this graded commutator satisfies the standard graded skew-symmetry and graded Jacobi identity:
\begin{equation}
\begin{gathered}
[\alpha, \beta] = -(-1)^{\lvert \alpha \rvert \lvert \beta \rvert} [\beta, \alpha], \\
(-1)^{\lvert \alpha \rvert \lvert \gamma \rvert}[\alpha, [\beta, \gamma]] + (-1)^{\lvert \beta \rvert \lvert \alpha \rvert}[\beta, [\gamma, \alpha]] + (-1)^{\lvert \gamma \rvert \lvert \beta \rvert}[\gamma, [\alpha, \beta]] = 0.
\end{gathered}
\end{equation}
We next note that the usual exterior derivative $\ud$ and the homological vector field $Q$ are graded derivations, with bigradings $(1,0)$ and $(0,1)$, respectively. The standard result for the exterior derivative acting on a product of forms also holds in this case for both derivations $\ud$ and $Q$, for example,
\begin{equation}
\ud (\alpha \wedge \beta) = \ud \alpha \wedge \beta + (-1)^{\deg(\alpha)} \alpha \wedge \ud \beta. 
\end{equation}
For the Hodge star operator $\ast$, we have, as usual,
\begin{equation}
{\ast}{\ast} \alpha = (-1)^{\deg (\alpha) } \alpha.
\end{equation}
Additionally, given a pure ghost form $\theta$, together with an arbitrary bigraded form $\alpha$, we have the following important property:
\begin{equation}
{\ast}[\theta,\alpha] = [\theta, {\ast}\alpha].
\end{equation}
Other properties to note are the trace formulae
\begin{equation}
\begin{gathered}
\tr \left(\alpha \wedge \beta \right) = (-1)^{\deg (\alpha) \deg (\beta)} \tr \left(\beta \wedge \alpha \right),  \\
\tr \left(\alpha \wedge [\beta,\gamma] \right) = \tr \left( [\alpha,\beta] \wedge \gamma \right),
\end{gathered}
\end{equation}
as well as the following integral formulae,
\begin{equation}
\begin{gathered}
\int_M \tr \left( \alpha \wedge \beta \right) = (-1)^{\deg(\alpha) + \deg(\beta)} \int_M \tr \left( {\ast}\alpha \wedge {\ast} \beta \right), \\
\int_M \tr \left(\alpha \wedge {\ast} [\beta,\gamma]  \right) = (-1)^{\lvert \alpha\rvert} \int_M \tr \left( {\ast}[{\ast}\alpha,\beta] \wedge {\ast}\gamma \right).
\end{gathered}
\end{equation}
The last two turn out to be particularly useful.  Our final formula refers to the way we compute functional derivatives. Such formula reads as follows:
\begin{equation}
\frac{\delta}{\delta \alpha} \int_{M} \tr \left( \alpha \wedge \beta \right) =  \beta. 
\end{equation}
Given all the rules listed above, the calculations performed over the course of section~\ref{sec:4} become straightforward, although some may still be tedious.


\section{The cohomology of the homological vector field $Q$}\label{app:C}
This short appendix computes the cohomology of the homological vector field $Q$ acting on functions on the graded manifold $T[4](\gfrak[1] \oplus \gfrak[2])$. To this end, it is convenient to write $\Wcal(\gfrak)$ for the algebra of such functions. Recall that, in terms of the Lie algebra coordinates $c$, $\phi$, $\sigma$ and $\xi$, viewed as the generators of $\Wcal(\gfrak)$, the action of $Q$ is given in \eqref{eq:4.2}. We will find it useful to replace the generators $\phi$ and $\xi$ by the generators $\bar{\phi}$ and $\bar{\xi}$ defined as
\begin{equation}
\begin{aligned}
\bar{\phi} &= \phi - \tfrac{1}{2}[c,c], \\
\bar{\xi} &= \xi - [c,\sigma].
\end{aligned}
\end{equation}
This change of variables has the key advantage that the action of $Q$ simplifies to
\begin{equation}
\begin{alignedat}{3}
Q c  &= -\bar{\phi}, &\quad Q \bar{\phi} &= 0, \\
Q \sigma &= -\bar{\xi}, &\quad  Q \bar{\xi} &= 0.
\end{alignedat}
\end{equation}
Once expressed in this form, it is now a simple matter to establish that $\Wcal(\gfrak)$, viewed as a cochain complex with differential $Q$, is acyclic. For a start, in degree $0$, the algebra $\Wcal(\gfrak)$ is simply $\RR$ and thus $H^0_Q(\Wcal(\gfrak)) = \RR$. For the other degrees, it is enough to find a cochain homotopy $K \colon \Wcal(\gfrak) \to \Wcal(\gfrak)$ of degree $-1$ such that for any homogeneous element $f \in \Wcal(\gfrak)$ of degree $n >0$,
\begin{equation}\label{eq:C.3}
(QK + KQ)f = n f.
\end{equation}
The reason this is sufficient is that, if $f$ is $Q$-closed, from \eqref{eq:C.3}, we have
\begin{equation}
f = \tfrac{1}{n} (QK + KQ)f  = \tfrac{1}{n} QKf,
\end{equation}
which confirms that $f$ is also $Q$-exact. To define the cochain homotopy $K$, we specify its action first on the generators $c$, $\bar{\phi}$, $\sigma$, and $\bar{\xi}$ as
\begin{equation}
\begin{alignedat}{3}
K c  &= 0, &\quad K \bar{\phi} &= -c, \\
K \sigma &= 0, &\quad  K \bar{\xi} &= -\sigma,
\end{alignedat}
\end{equation}
and extend it over all of $\Wcal(\gfrak)$ as a derivation. It is straightforward then to calculate that
\begin{equation}
\begin{alignedat}{3}
(QK + KQ)c  &= c, &\quad   (QK + KQ)\bar{\phi} &= \bar{\phi}, \\
(QK + KQ)\sigma &= \sigma, & \quad (QK + KQ)\bar{\xi} &= \bar{\xi}.
\end{alignedat}
\end{equation}
Suppose next that \eqref{eq:C.3} holds for a pair of homogeneous elements $f$ and $f'$ of $\Wcal(\gfrak)$ of degrees $n$ and $n'$, respectively. Then
\begin{equation}
(QK + KQ)(ff') = \left((QK + KQ) f \right) f' + f (QK + KQ)f'  = n ff' + f n' f' = (n + n') ff', 
\end{equation}
where in the first equality we used the fact that $QK + KQ$ is a derivation of degree $0$. So \eqref{eq:C.3} also holds for $ff'$. From this, we deduce that \eqref{eq:C.3} holds for all elements of $\Wcal(\gfrak)$, and therefore, we conclude that $H^n_Q (\Wcal(\gfrak)) = 0$ for $n > 0$.


\end{document}